\documentclass[letter, longauth]{aa}

\usepackage{graphicx}

\usepackage{amsmath}
\usepackage{amssymb}

\usepackage{lipsum}

\usepackage{natbib}
\usepackage{hyperref}

\newcommand{\RA}{\mathrm{RA}}
\newcommand{\DEC}{\mathrm{DEC}}

\begin{document}

\title{Direct confirmation of the radial-velocity planet $\beta$ Pic c}


\author{M.~Nowak\inst{\ref{cam},\ref{kavli}}
\and S.~Lacour\inst{\ref{lesia},\ref{esog}}
\and A.-M.~Lagrange\inst{\ref{ipag}}
\and P.~Rubini\inst{\ref{pixyl}}
\and J.~Wang\inst{\ref{caltech}}
\and T.~Stolker\inst{\ref{ethZ}}
\and R.~Abuter\inst{\ref{esog}}
\and A.~Amorim\inst{\ref{lisboa},\ref{centra}}
\and R.~Asensio-Torres\inst{\ref{mpia}}
\and M.~Bauböck\inst{\ref{mpe}}
\and M.~Benisty\inst{\ref{ipag}}
\and J.P.~Berger\inst{\ref{ipag}}
\and H.~Beust\inst{\ref{ipag}}
\and S.~Blunt\inst{\ref{caltech}}
\and A.~Boccaletti\inst{\ref{lesia}}
\and M.~Bonnefoy\inst{\ref{ipag}}
\and H.~Bonnet\inst{\ref{esog}}
\and W.~Brandner\inst{\ref{mpia}}
\and F.~Cantalloube\inst{\ref{mpia}}
\and B.~Charnay\inst{\ref{lesia}}
\and E.~Choquet\inst{\ref{lam}}
\and V.~Christiaens\inst{\ref{monash}}
\and Y.~Cl\'{e}net\inst{\ref{lesia}}
\and V.~Coud\'e~du~Foresto\inst{\ref{lesia}}
\and A.~Cridland\inst{\ref{leiden}}
\and P.T.~de~Zeeuw\inst{\ref{leiden},\ref{mpe}}
\and R.~Dembet\inst{\ref{esog}}
\and J.~Dexter\inst{\ref{mpe}}
\and A.~Drescher\inst{\ref{mpe}}
\and G.~Duvert\inst{\ref{ipag}}
\and A.~Eckart\inst{\ref{cologne},\ref{bonn}}
\and F.~Eisenhauer\inst{\ref{mpe}}
\and F.~Gao\inst{\ref{mpe}}
\and P.~Garcia\inst{\ref{centra},\ref{porto}}
\and R.~Garcia~Lopez\inst{\ref{dublin},\ref{mpia}}
\and T.~Gardner\inst{\ref{umich}}
\and E.~Gendron\inst{\ref{lesia}}
\and R.~Genzel\inst{\ref{mpe}}
\and S.~Gillessen\inst{\ref{mpe}}
\and J.~Girard\inst{\ref{stsci}}
\and X.~Haubois\inst{\ref{esoc}}
\and G.~Heißel\inst{\ref{lesia}}
\and T.~Henning\inst{\ref{mpia}}
\and S.~Hinkley\inst{\ref{exeter}}
\and S.~Hippler\inst{\ref{mpia}}
\and M.~Horrobin\inst{\ref{cologne}}
\and M.~Houllé\inst{\ref{lam}}
\and Z.~Hubert\inst{\ref{ipag}}
\and A.~Jiménez-Rosales\inst{\ref{mpe}}
\and L.~Jocou\inst{\ref{ipag}}
\and J.~Kammerer\inst{\ref{esog},\ref{australie}}
\and P.~Kervella\inst{\ref{lesia}}
\and M.~Keppler\inst{\ref{mpia}}
\and L.~Kreidberg\inst{\ref{mpia},\ref{harvard}}
\and M.~Kulikauskas\inst{\ref{pcc}}
\and V.~Lapeyr\`ere\inst{\ref{lesia}}
\and J.-B.~Le~Bouquin\inst{\ref{ipag}}
\and P.~L\'ena\inst{\ref{lesia}}
\and A.~Mérand\inst{\ref{esog}}
\and A.-L.~Maire\inst{\ref{liege},\ref{mpia}}
\and P.~Molli\`ere\inst{\ref{mpia}}
\and J.D.~Monnier\inst{\ref{umich}}
\and D.~Mouillet\inst{\ref{ipag}}
\and A.~Müller\inst{\ref{mpia}}
\and E.~Nasedkin\inst{\ref{mpia}}
\and T.~Ott\inst{\ref{mpe}}
\and G.~Otten\inst{\ref{lam}}
\and T.~Paumard\inst{\ref{lesia}}
\and C.~Paladini\inst{\ref{esoc}}
\and K.~Perraut\inst{\ref{ipag}}
\and G.~Perrin\inst{\ref{lesia}}
\and L.~Pueyo\inst{\ref{stsci}}
\and O.~Pfuhl\inst{\ref{esog}}
\and J.~Rameau\inst{\ref{ipag}}
\and L.~Rodet\inst{\ref{cornell}}
\and G.~Rodr\'iguez-Coira\inst{\ref{lesia}}
\and G.~Rousset\inst{\ref{lesia}}
\and S.~Scheithauer\inst{\ref{mpia}}
\and J.~Shangguan\inst{\ref{mpe}}
\and J.~Stadler\inst{\ref{mpe}}
\and O.~Straub\inst{\ref{mpe}}
\and C.~Straubmeier\inst{\ref{cologne}}
\and E.~Sturm\inst{\ref{mpe}}
\and L.J.~Tacconi\inst{\ref{mpe}}
\and E.F.~van~Dishoeck\inst{\ref{leiden},\ref{mpe}}
\and A.~Vigan\inst{\ref{lam}}
\and F.~Vincent\inst{\ref{lesia}}
\and S.D.~von~Fellenberg\inst{\ref{mpe}}
\and K.~Ward-Duong\inst{\ref{amherst}}
\and F.~Widmann\inst{\ref{mpe}}
\and E.~Wieprecht\inst{\ref{mpe}}
\and E.~Wiezorrek\inst{\ref{mpe}}
\and J.~Woillez\inst{\ref{esog}}
\and the GRAVITY Collaboration}

\institute{
  Institute of Astronomy, University of Cambridge, Madingley Road, Cambridge CB3 0HA, United Kingdom
  \label{cam}
\and
  Kavli Institute for Cosmology, University of Cambridge, Madingley Road, Cambridge CB3 0HA, United Kingdom
  \label{kavli}  
\and
  LESIA, Observatoire de Paris, Universit\'e PSL, CNRS, Sorbonne Universit\'e, Universit\'e de Paris, 5 place Jules Janssen, 92195 Meudon, France
  \label{lesia}
\and
  Universit\'e Grenoble Alpes, CNRS, IPAG, 38000 Grenoble, France
  \label{ipag}
\and
  Max Planck Institute for extraterrestrial Physics, Giessenbachstraße~1, 85748 Garching, Germany
  \label{mpe}
\and
  Max Planck Institute for Astronomy, K\"onigstuhl 17, 69117 Heidelberg, Germany
  \label{mpia}
\and
  European Southern Observatory, Karl-Schwarzschild-Straße 2, 85748 Garching, Germany
  \label{esog}
\and
  European Southern Observatory, Casilla 19001, Santiago 19, Chile
  \label{esoc}
\and
  Aix Marseille Univ, CNRS, CNES, LAM, Marseille, France
  \label{lam}
\and
  Department of Astronomy, California Institute of Technology, Pasadena, CA 91125, USA
  \label{caltech}
\and
  Space Telescope Science Institute, Baltimore, MD 21218, USA
  \label{stsci}
\and
  Astronomy Department, University of Michigan, Ann Arbor, MI 48109 USA
  \label{umich}
\and
  School of Physics and Astronomy, Monash University, Clayton, VIC 3800, Melbourne, Australia
  \label{monash}
\and
  Center for Astrophysics and Planetary Science, Department of Astronomy, Cornell University, Ithaca, NY 14853, USA
  \label{cornell}
\and
  1. Institute of Physics, University of Cologne, Z\"ulpicher Straße 77, 50937 Cologne, Germany
  \label{cologne}
\and
  Universidade de Lisboa - Faculdade de Ci\^encias, Campo Grande, 1749-016 Lisboa, Portugal
  \label{lisboa}
\and
  CENTRA - Centro de Astrof\'{\i}sica e Gravita\c c\~ao, IST, Universidade de Lisboa, 1049-001 Lisboa, Portugal
  \label{centra}
\and
  Leiden Observatory, Leiden University, P.O. Box 9513, 2300 RA Leiden, the Netherlands
  \label{leiden}
\and
  School of Physics, University College Dublin, Belfield, Dublin 4, Ireland
  \label{dublin}
\and
  Pasadena City College, Pasadena, CA 91106, USA
  \label{pcc}
\and
  Max Planck Institute for Radio Astronomy, Auf dem Hügel 69, 53121 Bonn, Germany
  \label{bonn}
\and
  STAR Institute/Université de Liège, Belgium
  \label{liege}
\and
  Center for Astrophysics | Harvard \& Smithsonian, Cambridge, MA 02138
  \label{harvard}
\and
  Pixyl, 5 Av. du Grand Sablon, 38700 La Tronche, France
  \label{pixyl}
\and
  Research School of Astronomy \& Astrophysics, Australian National University, ACT 2611, Australia
  \label{australie}
\and
  University of Exeter, Physics Building, Stocker Road, Exeter EX4 4QL, United Kingdom
  \label{exeter}
\and
  Institute for Particle Physics and Astrophysics, ETH Zurich, Wolfgang-Pauli-Strasse 27, 8093 Zurich, Switzerland
  \label{ethZ}
\and
  Universidade do Porto, Faculdade de Engenharia, Rua Dr. Roberto Frias, 4200-465 Porto, Portugal
  \label{porto}
\and
  Five College Astronomy Department, Amherst College, Amherst, MA 01002, USA
  \label{amherst}
}

\date{\today}
  \abstract
{Methods used to detect giant exoplanets can be broadly divided into two categories: indirect and direct. Indirect methods are more sensitive to planets with a small orbital period, whereas direct detection is more sensitive to planets orbiting at a large distance from their host star. 
This dichotomy makes it difficult to combine the two techniques on a single target at once.}
{Simultaneous measurements made by direct and indirect techniques offer the possibility of determining the mass and luminosity of planets and a method of testing formation models. Here, we aim to show how long-baseline interferometric observations guided by radial-velocity can be used in such a way.} 
{We observed the recently-discovered giant planet $\beta$ Pictoris c with GRAVITY, mounted on the Very Large Telescope Interferometer (VLTI).}
{This study constitutes the first direct confirmation of a planet discovered through radial velocity. We find that the planet has a temperature of
$T = 1250\pm50$\,K and a dynamical mass of $M = 8.2\pm0.8\,M_{\rm Jup}$. At $18.5\pm2.5$\,Myr, this puts $\beta$ Pic c close to a 'hot start' track, which is usually associated with formation via disk instability. Conversely, the planet orbits at a distance of
2.7\,au, which is too close for disk instability to occur. The low apparent magnitude ($M_{\rm K} = 14.3 \pm 0.1$) favours a core  accretion scenario.}
{We suggest that this apparent contradiction is a sign of hot core accretion, for example, due to the mass of the planetary core or the existence of a high-temperature accretion shock during formation.}


   \maketitle

\section{Introduction}

\begin{table*}
  \begin{center}
    \begin{tabular}{cccccc}
      \hline
      \hline
      Date & UT Time &Nexp / NDIT / DIT & airmass & tau$_0$ & seeing \\
      \hline
      2020-02-10 & 02:32:52 - 04:01:17 & 11 / 32 / 10\,s & 1.16-1.36 & 6-18\,ms & 0.5-0.9" \\
      2020-02-12 & 00:55:05 - 02:05:29 &11 / 32 / 10\,s  & 1.12-1.15 & 12-23\,ms & 0.4-0.7"\\
      2020-03-08 & 00:15:44 - 01:41:47 & 12 / 32 / 10\,s & 1.13-1.28 & 6-12\,ms & 0.5-0.9" \\
      \hline
      \hline
    \end{tabular}
    \caption{Observing log.}
    \label{tab:log}
  \end{center}
\end{table*}

\begin{figure*}
  \begin{center}
    \begin{minipage}{0.32\linewidth}{}
      \hspace{0.12\linewidth} 9 February 2020 \\[0.2cm]
        \includegraphics[width=\linewidth, clip=True, trim=2cm 1.45cm 1cm 1.55cm]{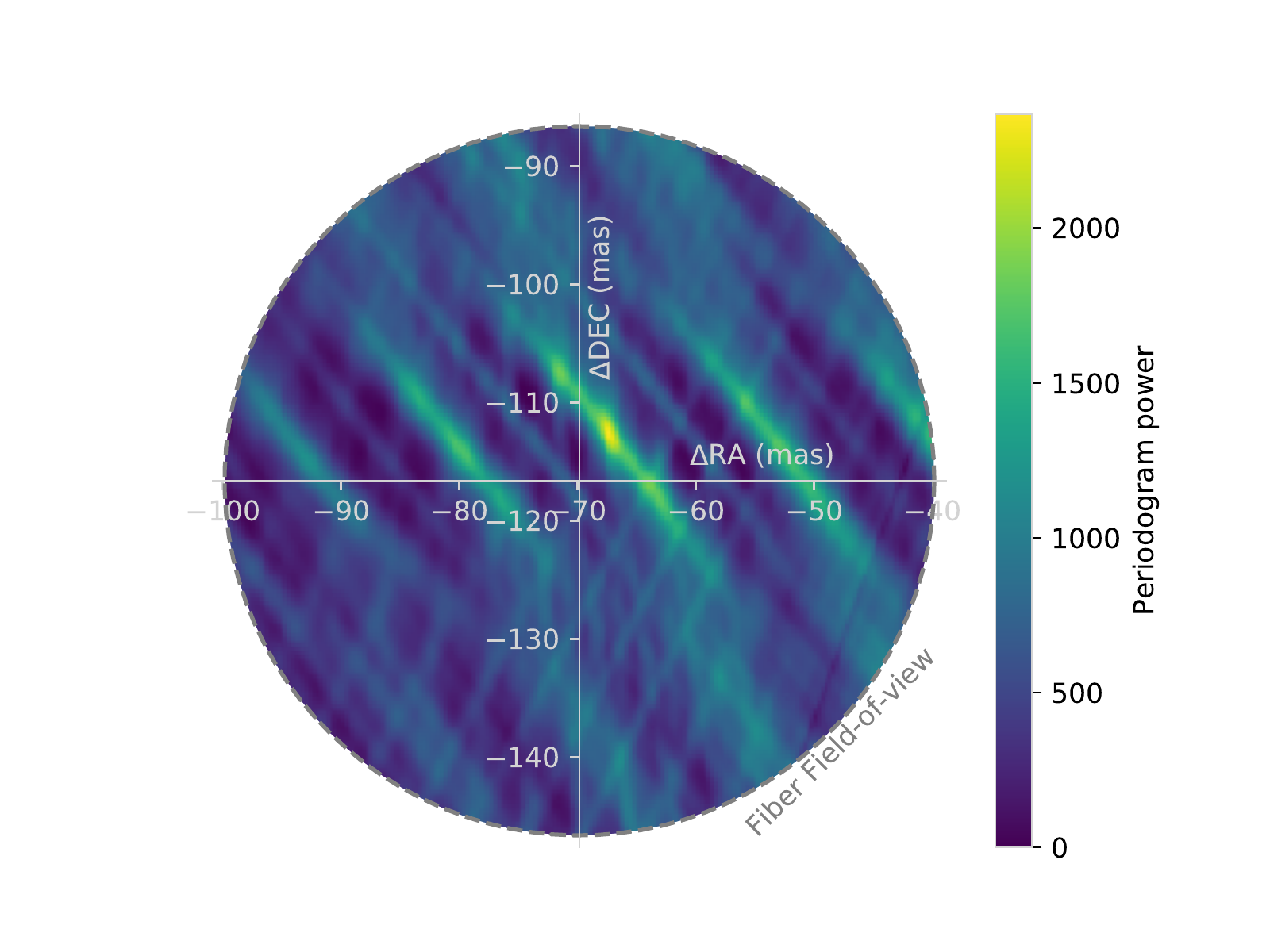}
    \end{minipage}
    \begin{minipage}{0.32\linewidth}{}
      \hspace{0.12\linewidth} 11 February 2020 \\[0.2cm]
        \includegraphics[width=\linewidth, clip=True, trim=2cm 1.45cm 1cm 1.55cm]{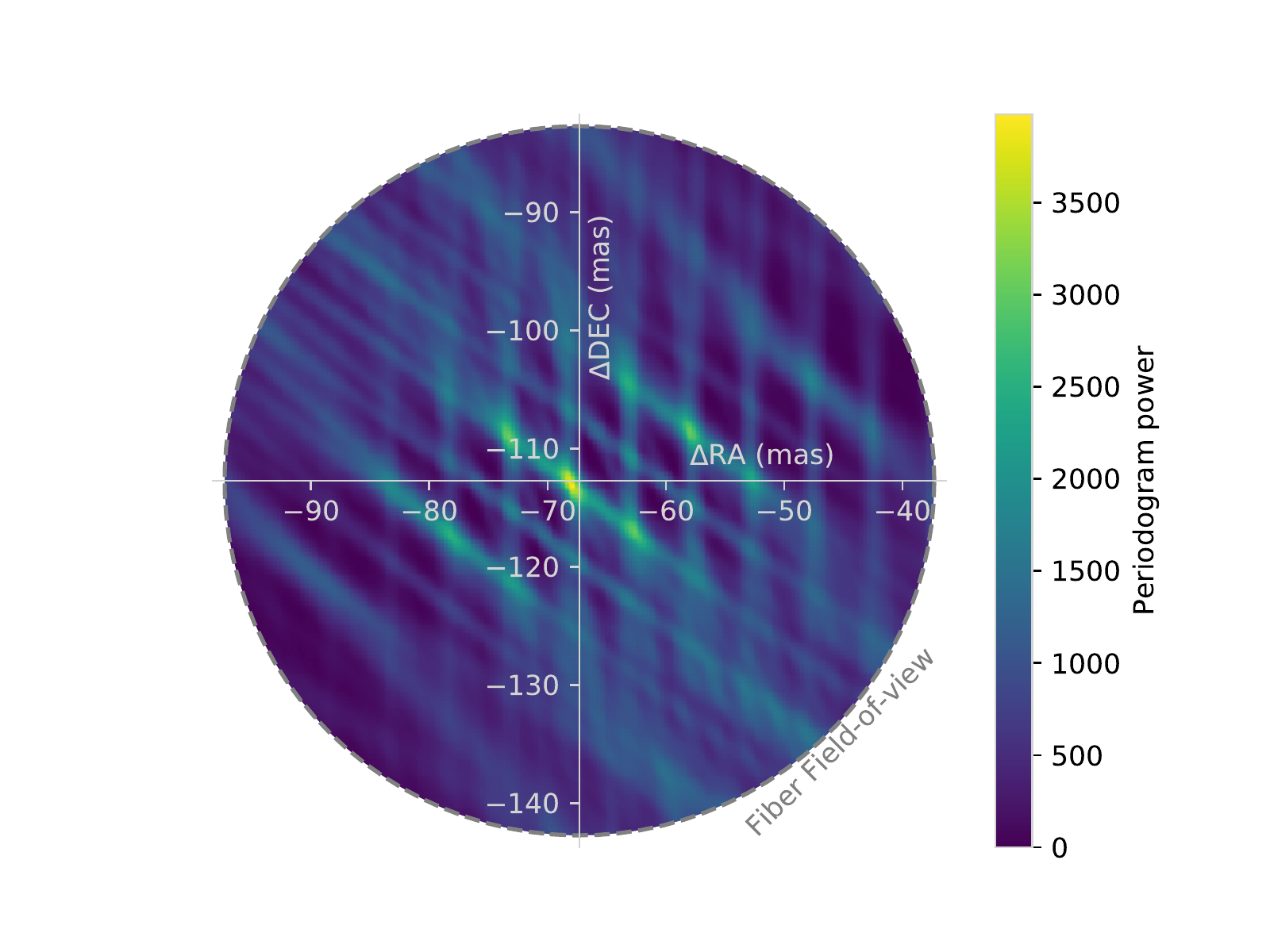}
    \end{minipage}
    \begin{minipage}{0.32\linewidth}{}
      \hspace{0.15\linewidth} 7 March 2020 \\[0.2cm]
        \includegraphics[width=\linewidth, clip=True, trim=2cm 1.45cm 1cm 1.55cm]{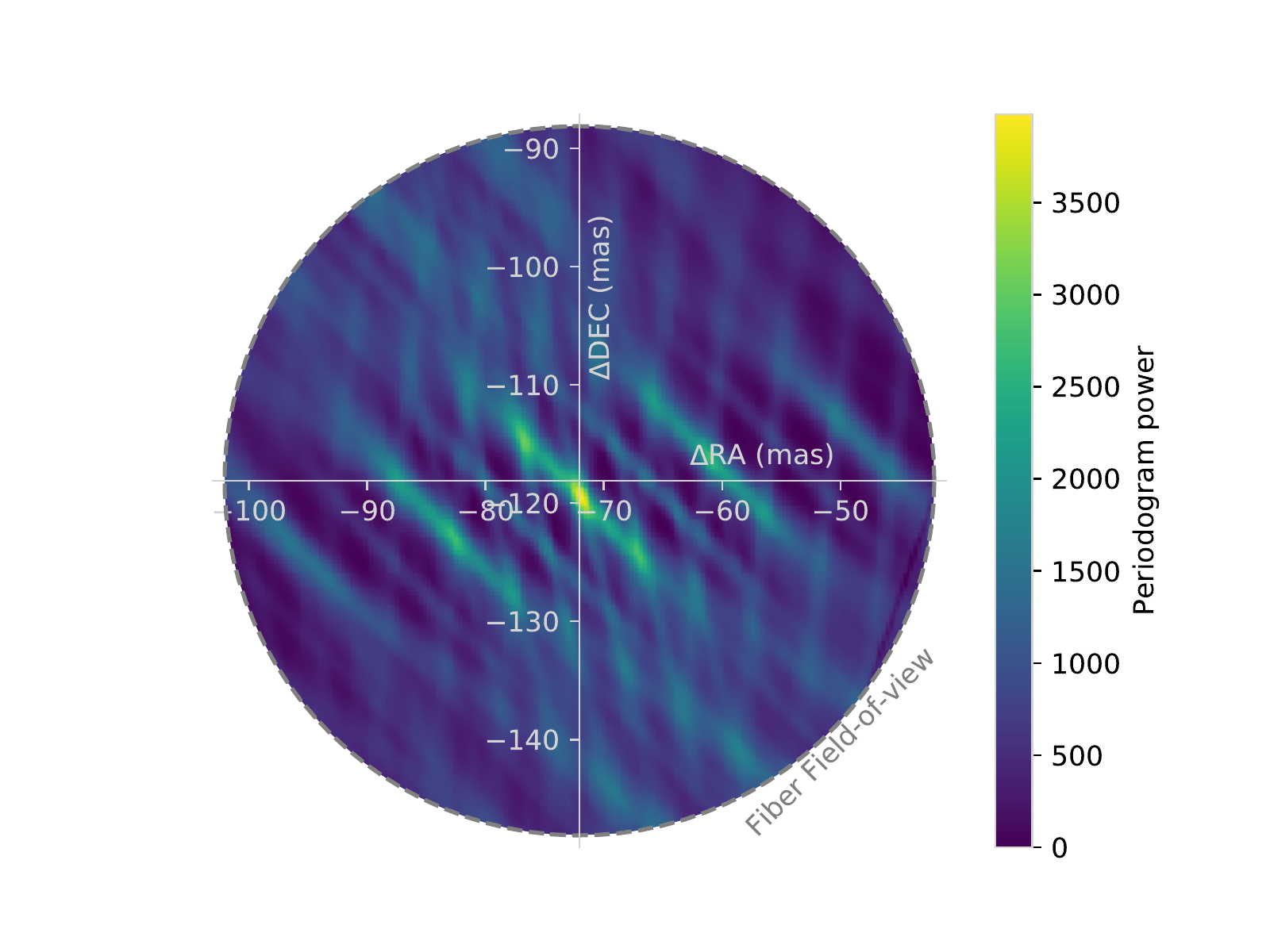}
     \end{minipage}
    \caption{Detection of $\beta$ Pictoris c. The three panels show the periodogram power maps calculated over the fiber field-of-view for each of the three observation epochs and following the subtraction of stellar residuals. The presence of a peak in the power maps indicates a point-like source in the field-of-view of the fiber, with side lobes that are characteristic of the interferometric nature of the observations.}
    \label{fig:chi2Maps}
  \end{center}
\end{figure*}

Giant planets are born in circumstellar disks from the material that remains after the formation of stars. The processes by which they form remain unclear and two schools of thought are currently competing to formulate a valid explanation, based on  two scenarios: 1) disk instability, which states that planets form through the collapse and fragmentation of the circumstellar disk \citep{Boss1997, Cameron1978}, and 2) core accretion, which states that a planetary core forms first through the slow accretion of solid material and later captures a massive gaseous atmosphere \citep{Pollack1996,2007prpl.conf..591L}. 


These two scenarios were initially thought to lead to very different planetary masses and luminosities, with planets formed by gravitational instability being much hotter and having higher post-formation luminosity and entropy than planets formed by core accretion \citep{Marley2007}. With such a difference in the post-formation entropy, the two scenarios could be distinguished using mass and luminosity measurements of young giant planets. However, several authors have since shown that so-called `hot start' planets are not incompatible with a formation model of core accretion, provided that the physics of the accretion shock \citep{Marleau2017, Marleau2019} and the core-mass effect {(a self-amplifying process by which a small increase of the planetary core mass weakens the accretion shock during formation and leads to higher post-formation luminosity)} is properly accounted for \citep{Mordasini2013, Mordasini2017}. In this situation, mass and luminosity measurements are of prime interest as they hold important information regarding the physics of the formation processes.

The most efficient method used thus far to determine the masses of giant planets is radial velocity measurements of the star. This method has one significant drawback: it is only sensitive to the product, $m\sin(i)$, where $m$ is the mass of the planet and $i$ its orbital inclination. In parallel, the most efficient method used to measure the luminosities of giant planets is direct imaging using dedicated high-contrast instruments. Since direct imaging also provides a way to estimate the orbital inclination {when the period allows for a significant coverage of the orbit}, the combination of radial velocity and direct imaging can, in principle, break the $m\sin(i)$ degeneracy and enable an accurate measurement of masses and luminosities. {But therein lies the rub: while radial-velocity is sensitive to planets orbiting close-in (typically $<1\,\mathrm{au}$) around old ($>1\,\mathrm{Gyr}$) stars, direct imaging is sensitive to planets orbiting at much larger separations ($\ge{}10\,\mathrm{au}$) around younger stars ($<100\,\mathrm{Myr}$). Thus, these techniques are not easily combined for an analysis of a single object.}

Significant efforts have been made over the past few years to extend the radial-velocity method to longer periods and younger stars. This is illustrated by the recent detection of $\beta$ Pic c \citep{Lagrange2019}. The recent characterisation of HR~8799~e also demonstrated the potential of interferometric techniques to determine orbits, luminosities \citep{GRAVITYCollaboration2019}, and atmospheric properties \citep{Molliere2020} of imaged extra-solar planets with short periods.
In this paper, we attempt to bridge the gap between the two techniques by reporting
on the direct confirmation of a planet discovered by radial velocity: $\beta$
Pic c.

\section{Observations and detection of $\beta$ Pic c}
\label{sec:obs}

\subsection{Observations}

We observed $\beta$ Pic c during the night of the $9^{\rm th}$ and $11^{\rm th}$ of February 2020, as well as that of the $7^{\rm th}$ of March 2020 with the Very Large Telescope Interferometer (VLTI), using the four 8.2~m Unit Telescopes (UTs), and the GRAVITY instrument \citep{GRAVITYCollaboration2017}. The first detection was obtained from the allocated time of the ExoGRAVITY large program (PI Lacour, ID 1104.C-0651). A confirmation was obtained two days later courtesy of the active galactic nucleus (AGN) large programme (PI Sturm, ID 1103.B-0626). The final dataset was obtained with the Director's Discretionary Time (DDT; ID 2104.C-5046). Each night, the atmospheric conditions ranged from good {(0.9")} to excellent {(0.4")}. The observing log is presented in Table~\ref{tab:log}, and the observing strategy is described in Appendix~\ref{app:obs}. In total, three~hours of integration time were obtained in K-band ($2.2~\mu\mathrm{m}$) at medium resolution (R=500).

\subsection{Data reduction and detection}

\begin{figure*}
  \begin{center}
    \includegraphics[width=0.9\linewidth]{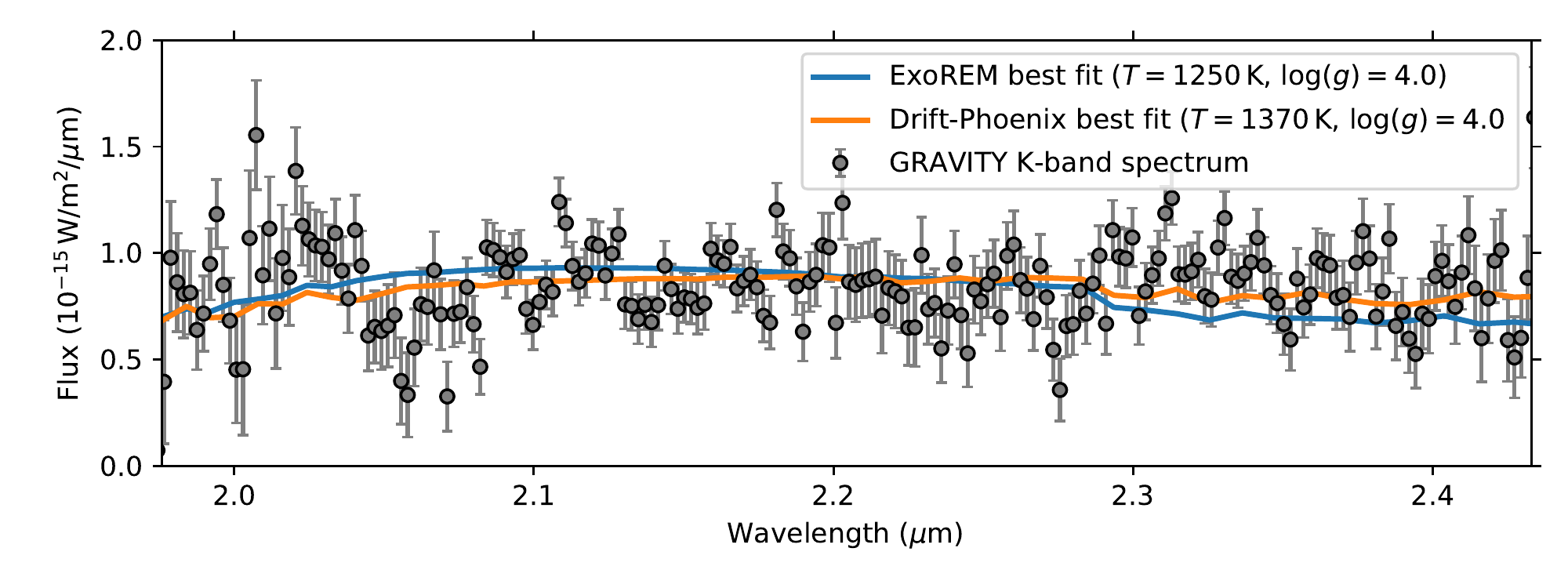}        \caption{K-Band spectrum of $\beta$ Pic c, with the best fit obtained with the Exo-REM and Drift-Phoenix grids overplotted (see Section~\ref{sec:physics}).}
    \label{fig:spectrum}
  \end{center}
\end{figure*}

\begin{table}
  \begin{center}
    \begin{tabular}{c c c c c c}
      \hline
      \hline
      MJD & $\Delta\RA$ & $\Delta\DEC$ & $\sigma_{\Delta\RA}$ & $\sigma_{\Delta\DEC}$ & $\rho$ \\
      (days) & (mas) & (mas) & (mas) & (mas) & - \\
      \hline
      58889.140 & -67.35 & -112.60 & 0.16 & 0.27 & -0.71 \\
      58891.065 & -67.70 & -113.18 & 0.09 & 0.18 & -0.57 \\
      58916.043 & -71.89 & -119.60 & 0.07 & 0.13 & -0.43 \\            
      \hline
      \hline
    \end{tabular}
    \caption{Relative astrometry of $\beta$ Pic c extracted from our VLTI/GRAVITY observations. Due to the interferometric nature of the observations, a correlation coefficient $\rho$ is required to properly describe the confidence intervals, which are not aligned on the sky coordinates. The covariance matrix can be reconstructed using ${\sigma_{\Delta\RA}}^2$ and ${\sigma_{\Delta\DEC}}^2$ on the diagonal, and $\rho\sigma_{\Delta\RA}\sigma_{\Delta\DEC}$ off-diagonal.}
    \label{tab:astrometry}
  \end{center}
\end{table}

The data reduction used for $\beta$ Pic c follows what has been developed for $\beta$ Pic b \citep{GRAVITYCollaboration2020} and is detailed in Appendix~\ref{app:datared}.

An initial data reduction using the GRAVITY pipeline \citep{Lapeyrere2014} gives the {interferometric visibilities obtained on the planet, and on the star, phase-referenced with the metrology system (i.e. in which the atmospheric and instrumental phase distortions are reduced to an unknown constant).}
The data reduction then proceeds in two steps, namely an initial extraction of the astrometry followed by the extraction of the spectrum. {Since the planet and the star are observed successively with the instrument, the data reduction yields a contrast spectrum, defined as the ratio of the planet spectrum to the star spectrum. This observable is more robust to variations of the instrument and atmospheric transmission.} The overall process yields a periodogram power map in the field-of-view of the science fiber (see Figure~\ref{fig:chi2Maps}), and a contrast spectrum. The astrometry (see Table~\ref{tab:astrometry}) is extracted from the periodogram power maps by taking the position of the maximum of the periodogram power, $z$, and error bars are obtained by breaking each night into individual exposures {(11 to 12 per night, see Table~\ref{tab:log})}, so as to estimate the effective standard-deviation from the data themselves.

{The planet spectrum is obtained by combining the three K-band contrast spectra obtained at each epoch (no evidence of variability was detected) and multiplying them by a NextGen stellar model \citep{Hauschildt1999} corresponding to the known stellar parameters (temperature $T=8000~\mathrm{K}$, surface gravity $\log(g) = 4.0\,\mathrm{dex}$, \citeauthor{Zorec2012} \citeyear{Zorec2012}) and a solar metallicity, scaled to an ESO K-band magnitude of 3.495 \citep{Bliek1996}. We note that at the resolution of GRAVITY in the K-band, the temperature, surface gravity, and metallicity of the star all have only a marginal impact on the final spectrum. The resulting flux calibrated K-band spectrum is presented in Figure~\ref{fig:spectrum}.}

\section{The orbit of $\beta$ Pic c}
\label{sec:orb}

\begin{table*}
  \begin{center}
    \begin{tabular}{lcccl}
      \hline
      \hline
      Parameter & Prior distribution &  \multicolumn{2}{c}{Posteriors (1$\sigma$)} &   Unit \\
      \hline
      \hline
      Star & & \multicolumn{2}{c}{$\beta$ Pictoris}  & \\
      \hline
 Stellar mass & Gaussian ($1.77\pm0.02)$  &\multicolumn{2}{c}{$1.82 \pm  0.03$}  &  M$_{\odot}$\\
 Parallax & Uniform & \multicolumn{2}{c}{$51.42 \pm  0.11$} &   mas \\
 proper motion & Uniform & \multicolumn{2}{c}{ra=$4.88 \pm 0 .02$, dec=$83.96 \pm 0 .02$} &   mas/yr \\
 RV v$_0$ & Uniform & \multicolumn{2}{c}{$-23.09 \pm  8.44$} &   m/s \\
 RV jitter & Uniform & \multicolumn{2}{c}{$31.21 \pm  8.09$} &   m/s \\
      \hline
      \hline
      Planets  & & $\beta$ Pictoris b & $\beta$ Pictoris c &\\
      \hline
 Semi-major axis &  Log Uniform & $9.90 \pm  0.05$ &  $2.72 \pm  0.02$ &au \\
 Eccentricity & Uniform &  $0.10 \pm  0.01$ &   $0.37 \pm  0.12$   &\\
 Inclination & Sin Uniform & $88.99 \pm  0.01$ &  $89.17 \pm  0.50$ & deg \\
 PA of ascending node & Uniform & $31.82 \pm  0.02$  & $30.98 \pm  0.12$  & deg \\
 Argument of periastron & Uniform & $196.9 \pm  3.5$ &  $66.2 \pm  2.5$ & deg \\
 Epoch of periastron & Uniform &  $0.72 \pm  0.01$ &  $ 0.83 \pm  0.02$ & \\
 Planet mass & Gaussian ($15.4\pm3.0)$  & $9.0 \pm  1.6$ &   & M$_{\rm jup}$ \\
   & Log Uniform &  &  $ 8.2 \pm  0.8$ & M$_{\rm jup}$ \\
   \hline
   \hline
 Period &  --- & $23.28 \pm  0.46$ & $3.37 \pm  0.04$ & years \\
 $\Delta$m$_{\rm K}$ & --- & $8.9\pm0.1$ & $10.8\pm0.1$ & mag \\ 
      \hline
      \hline
    \end{tabular}
    \caption{Posteriors of the MCMC analysis of the $\beta$ Pictoris system. Orbital elements are given in Jacobi coordinates. 
 The ascending node is defined to be the node where the motion of the companion is directed away from the Sun.
 The epoch of periastron is given by  \texttt{orbitize!} as a fraction of the period from a given date (here, 58889 MJD). The difference between the $\beta$ Pic b mass from the prior and posterior indicates a tension between the dynamical data and spectral analysis. The periods are calculated from the posteriors. The K band delta magnitudes are obtained from present and published \citep{GRAVITYCollaboration2020} GRAVITY spectra. }
    \label{tab:fit_parameters}
  \end{center}
\end{table*}

\begin{figure}
  \begin{center}
    \includegraphics[width=\linewidth]{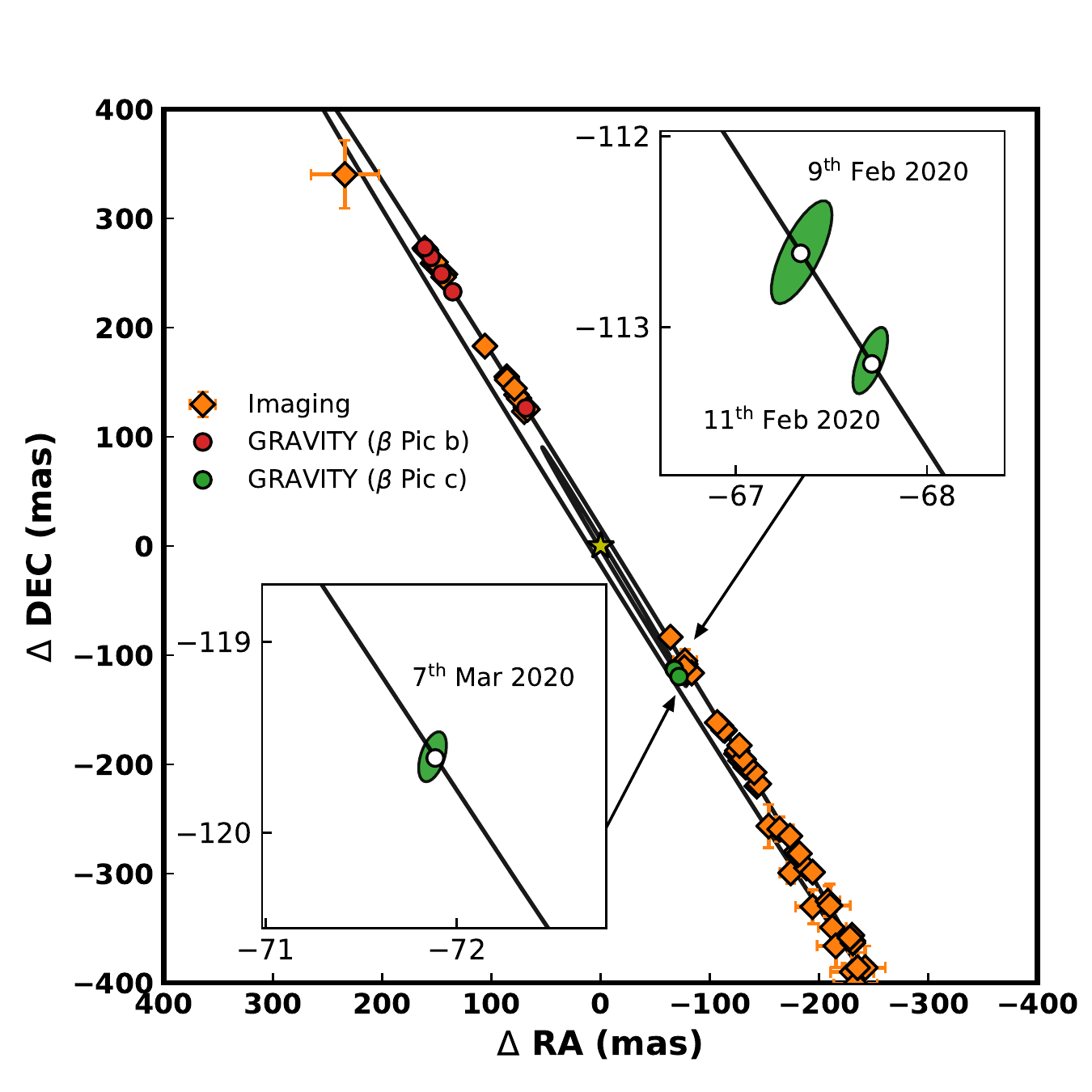}
    \caption{Motion of planets b and c around  $\beta$ Pictoris, as observed by GRAVITY (respectively red and green dots) and direct imaging (orange dots). The inset panels are zooms on the GRAVITY observations of  $\beta$ Pic c. Within the insets, the green ellipses correspond to the $1\,\sigma$ error intervals, typically between 100  to 200 $\mu$as, elongated because of the shape of the interferometric array.  The white dots 
    are the retrieved position at the time of observation from the posteriors in Table~\ref{tab:fit_parameters}.}
    \label{fig:motion}
  \end{center}
\end{figure}

{
The relative astrometry is plotted in Figure~\ref{fig:motion}. 
We used a Markov Chain Monte Carlo (MCMC) sampler to determine the orbital parameters of both planets, as well as the mass of each object. The MCMC analysis was done using the  \texttt{orbitize!}\footnote{documentation available at \url{http://orbitize.info}} \citep{Blunt2020} software. For the relative astrometry,  we included all previous direct imaging and GRAVITY data. The NACO, SPHERE, and GRAVITY observations of $\beta$ Pic  b are from \citet{Lagrange2020} and reference therein. The 
existing GPI data are summarized by \citet{Nielsen2020}.
Of course, we added the three GRAVITY detections of $\beta$ Pic c from Table~\ref{tab:astrometry}.
For the absolute astrometry of the star, we used the Hipparcos Intermediate Astrometric Data \citep{Leeuwen2007} and the DR2 position \citep{GAIACollaboration2018}. Lastly, for the radial velocity, we used the 12 years of HARPS measurements \citep{Lagrange2019,Lagrange2020}.
}

{
The priors of the MCMC analysis are listed in Table~\ref{tab:fit_parameters}.
Most of the priors are uniform or pseudo-uniform distributions. We have made sure that all uniform distributions have limits many sigma outside the distributions of the posteriors.
We used two Gaussian priors to account for independent knowledge. Firstly, the knowledge of the mass of b that comes from the spectrum
of the atmosphere \citep{GRAVITYCollaboration2020}: $15.4\pm 3.0\,{\rm M}_{\rm Jup}$. Secondly, an estimation of the stellar mass of  $1.75\pm0.05$\,M$_{\odot}$ \citep{Kraus2020}.
}

The semi-major axis of $\beta$ Pic c's orbit is  $2.72\pm0.02\,$au, with an orbital period of $\beta$ Pic c is $3.37\pm0.04$ year. The planet is currently moving toward us, with a passage of the Hill sphere between Earth and the star in mid-August 2021 ($\pm1.5$\,month).{ This means that the transit of circumplanetary material could be seen at that time. }
However, no sign of a photometric event was detected at the time of the previous conjunction, which was between February-June 2018 \citep{Zwintz2019}.

The position angles of the ascending nodes of b and c are  $31.82\pm0.02\,^\circ$ and $30.98\pm0.12\,^\circ$, respectively. Their inclinations are $88.99\pm0.01^\circ$ and $89.17\pm0.50\,^\circ$. This means that the two planets are co-planar to within less than a degree, perpendicular from the angular momentum vector of the star \citep{Kraus2020}.
The eccentricity of c from radial velocity alone is $0.29\pm0.05$ \citep{Lagrange2020}. {By including the new relative astrometry, we find  a higher value, which is consistent within the error bars, of $0.37\pm0.12$.} In combination with the eccentricity of b ($0.10\pm0.01$), the system has  a significant reservoir of eccentricity.

{
Using the criterion for stability from \cite{2015ApJ...808..120P}, we establish that the system is likely to be stable for at least 100 Myr. Using an N-body simulation \citep[SWIFT HJS]{2003A&A...400.1129B}, we confirm this stability for 10 million years for the peak of the orbital element posteriors. Due to the high mass ratios and eccentricities, the periodic secular variations of the inclinations and eccentricities are not negligible on this timescale. The periods associated with the eccentricities are around 50 000 year and can trigger variations of 0.2 in the eccentricities of both bodies, while the period associated with the relative inclination is around 10 000 year with variations by 20\%. Moreover, instabilities that could produce eccentricity may still happen on longer time scales, within a small island of the parameter space or by interaction with smaller planetary
bodies. }
 One possibility, for example, is that the planetary system is undergoing dynamical perturbations from planetesimals, as our own solar system may well have undergone in the past \citep[the Nice model,][]{Nesvorny2018}.

With a knowledge of the inclination, the mass of $\beta$~Pic~c can be determined from the radial velocity data. In addition, the mass of c can theoretically be obtained purely from the astrometry of b.
This can be explained by the fact that the orbital motion of c affects the position of the central star, which sets the orbital period of c on the distance between b and the star.
The amplitude of the effect is twice $2.72\times( {\rm M}_{\rm c}/{\rm M}_{\rm star})$, equivalent to $\approx 1\,200\,\mu$as. This is many times the GRAVITY error bar, so this effect must be accounted for in order to estimate the orbital parameters of b (see Appendix~\ref{app:orbit}). In a few years, once the GRAVITY data have covered a full orbital period of c, this effect will give an independent  dynamical mass  measurement of c. 
{
For now, the combination of RV and astrometry gives a robust constrain of $M_{c}=8.2\pm0.8\,M_{\rm Jup}$, compatible with  the estimations by \citet{Lagrange2020}.
 We observed a mass for $\beta$ Pic b ($M_b=9.0\pm1.6\,M_{\rm Jup}$) that is 2$\sigma$ from the mean of the Gaussian prior. If we remove the prior from the analysis, we find a  significantly lower mass: $M_b=5.6\pm1.5\,M_{\rm Jup}$. 
 Therefore, the radial velocities bias the data toward a lighter planet,
 discrepant with other analyses using only absolute astrometry:  $11\pm2\,M_{\rm Jup}$ in \citet{Snellen2018}, $13.7^{+6}_{-5}\,M_{\rm Jup}$ in \citet{Kervella2019} and $12.8^{+5.3}_{-3.2}\,M_{\rm Jup}$ in \citet{Nielsen2020}. These three results, however, should be taken with caution; the analyses were carried out based on a single planet. In our MCMC analysis with two planets, we could not reproduce these high masses. This is because the planet c  is responsible for  a large quantity of the proper motion observed with \emph{Hipparcos},  converging to a low mass for $\beta$ Pic b. 
 }

\section{Physical characteristics of $\beta$ Pic c}
\label{sec:physics}
\subsection{Atmospheric modelling}

The spectrum extracted from our GRAVITY observations can be used in conjunction with atmospheric models to constrain some of the main characteristics of the planet; 
the overall K-band spectral shape is driven mostly by the temperature and, to a lesser extent, by the surface gravity and metallicity. Since the GRAVITY spectrum is flux-calibrated, the overall flux level also constrains the radius of the planet.

Using the Exo-REM atmospheric models \citep{Charnay2018} and the \texttt{species}\footnote{available at \url{https://species.readthedocs.io}} atmospheric characterization toolkit \citep{Stolker2020}, we find a temperature of $T=1250\pm{}50\,\mathrm{K}$, a surface gravity of $\log(g) = 3.85^{+0.15}_{-0.25}$, and a radius of $1.2\pm{}0.1\,R_\mathrm{Jup}$.

{The Drift-Phoenix model \citep{Woitke2003, Helling2006, Helling2008} gives a slightly higher temperature ($1370\pm{}50\,K$) and a smaller radius ($1.05\pm{}0.1\,R_\mathrm{Jup}$). The surface-gravity is similar to the one obtained with Exo-REM with larger errors ($\log(g) = 4.1\pm{}0.5\,\mathrm{dex}$).}

{The best fits obtained with Exo-REM and Drift-Phoenix are overplotted to the GRAVITY spectrum in Figure~\ref{fig:spectrum}. We note that at this temperature and surface gravity, both the Exo-REM and Drift-Phoenix models are relatively flat and do not show any strong molecular feature. This is also the case of the GRAVITY spectrum, which remains mostly flat over the entire K-band. The only apparent spectral feature is located at $\simeq{}2.075\,\mu\mathrm{m}$. It is not reproduced by the models, and is most probably of telluric origin.}

Interestingly, the temperature and surface gravity derived from the atmospheric models (which are agnostic regarding the formation of the planet) are relatively close to the predictions of the `hot start' AMES-Cond evolutionary tracks \citep{Baraffe2003}. For a mass of $8.2\pm{}0.8\,M_\mathrm{Jup}$ and an age of $18.5\pm{}2.5\,\mathrm{Myr}$ \citep{MiretRoig2020}, the AMES-Cond tracks predict a temperature of $T=1340\pm{}160\,K$ and a surface gravity of $\log(g)=4.05\pm{}0.05$.


\subsection{Mass/luminosity and formation of $\beta$ Pic c}

\begin{figure*}
  \begin{center}
    \includegraphics[width=0.49\linewidth]{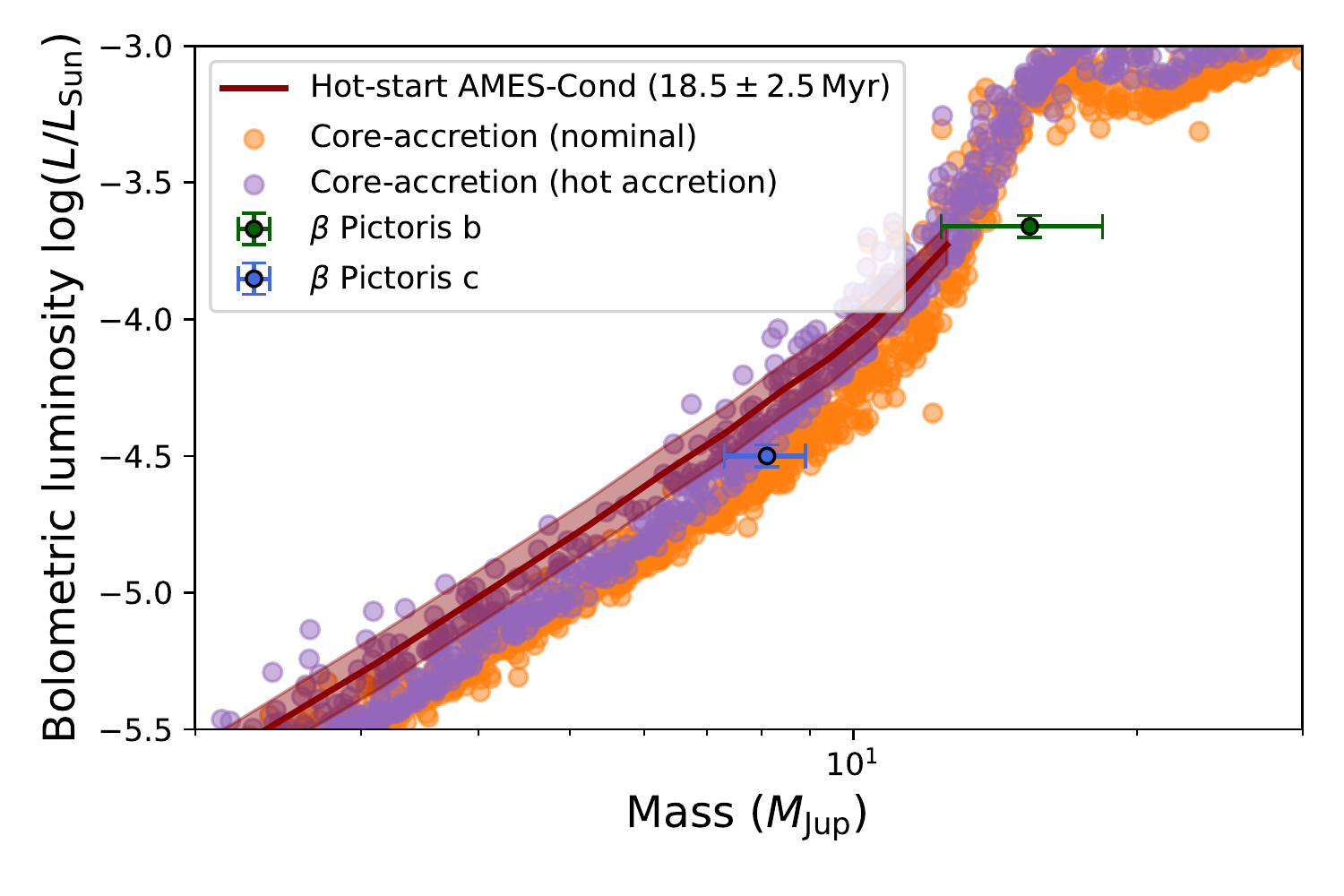}    \includegraphics[width=0.49\linewidth]{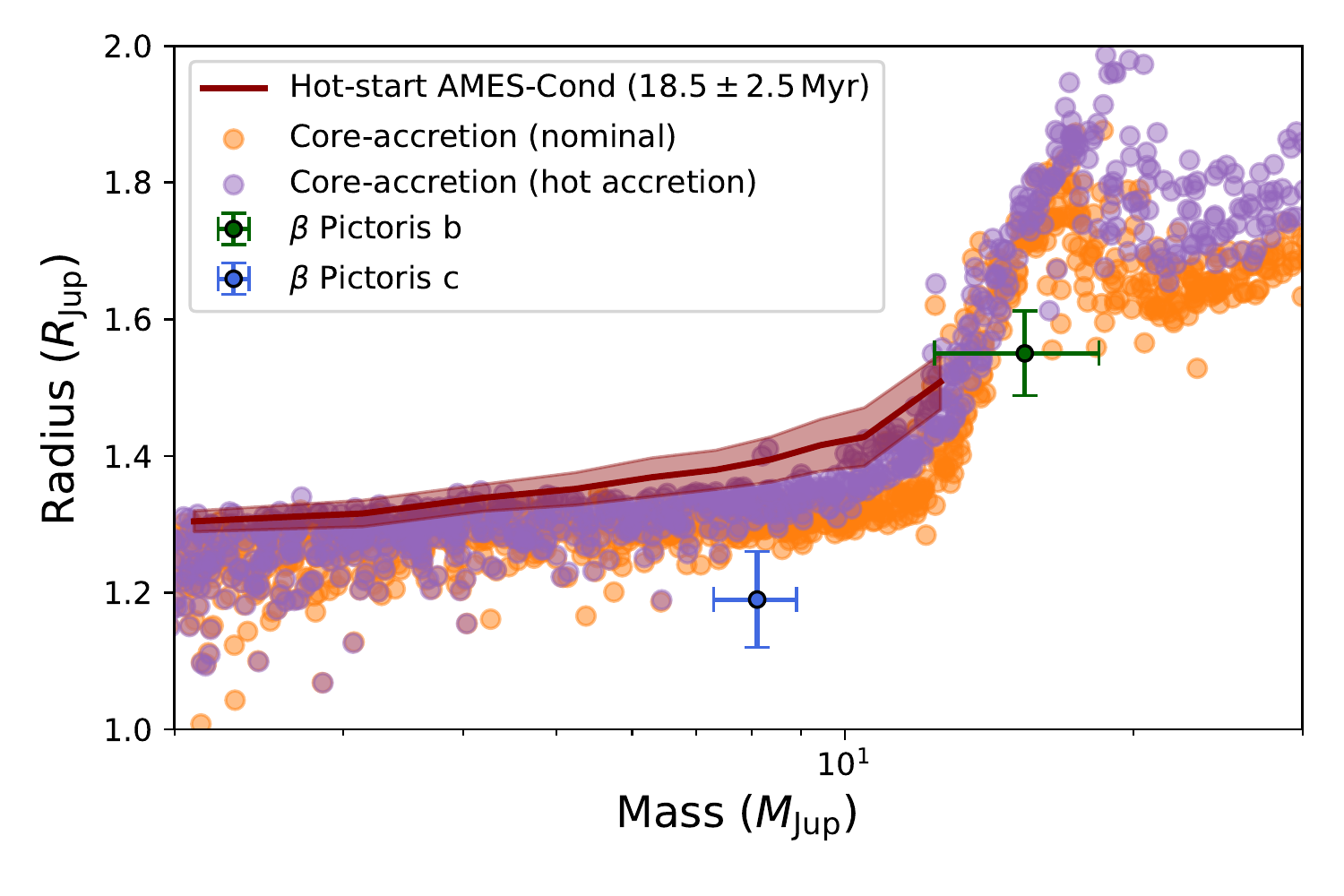}
    \caption{Position of $\beta$ Pic b \& c in mass/luminosity and mass/radius. The two panels give mass/radius(left-hand side) and mass/radius (right-hand side) diagrams showing the positions of the two known giant planets of the $\beta$ Pic system and the predictions of planet formation models. The predictions of the AMES-Cond `hot start' model, which extends to $\simeq{}14~M_\mathrm{Jup}$, are depicted as a red line (with the shaded area corresponding to the uncertainty on the age). Two synthetic populations generated by core accretion are also shown: CB753, corresponding to core accretion with cold nominal accretion and CB752, with hot accretion. These two populations are valid for 20 Myr and take into account the core-mass effect \citep{Mordasini2013}.}
    \label{fig:mass_luminosity}
  \end{center}
\end{figure*}

Such hot young giant planets $(>1000\,K)$ have historically been associated with formation through disk instability \citep{Marley2007}. However, disk fragmentation is unlikely to occur within the inner few au \citep{Rafikov2005} and, in that respect, the existence of a planet as massive as $\beta$~Pic~c at only $2.7\,\mathrm{au}$ poses a real challenge.

One possibility is that the planet formed in the outer $\beta$ Pic system ($\gtrapprox 30$\,au) and later underwent an inward type II migration that brought it to its current orbit. From a dynamical standpoint, however, the plausibility of such a scenario given the existence of a second giant planet in the system remains to be proven.

An alternative to the formation of $\beta$ Pic c by disk instability is core-accretion. In recent years, a number of authors have shown that it is possible to obtain hot and young giant planets with such a scenario if the core is sufficiently massive \citep{Mordasini2013}, if the accretion shock is relatively inefficient at cooling the incoming gas \citep{Mordasini2017, Marleau2014}, or if the shock dissipates sufficient heat into the in-falling gas \citep{Marleau2017, Marleau2019}.

A possible argument in favor of a core accretion scenario comes from the marginal discrepancy in luminosity between the AMES-Cond `hot-start' model and our observations. The spectrum given in Figure~\ref{fig:spectrum} is flux-calibrated and can be integrated over the K-band to give the apparent magnitude of the planet. {To calculate this magnitude, we first interpolate the ESO K-band transmission curve~\citep{Bliek1996} over the GRAVITY wavelength bins to obtain a transmission vector, $T$. Then, denoting $S$ as the flux-calibrated planet spectrum and $W$ as the associated covariance matrix, the total flux in the ESO K-band is: $F=T^TS$. The error bar is given by: $\Delta{}F = TWT^T$. The fluxes are then converted to magnitudes using the proper zero-points \citep{Bliek1996}. This gives a K-band magnitude for $\beta$~Pic~c of $m_\mathrm{K} = 14.3\pm{0.1}$ ($\Delta m_\mathrm{K} = 10.8\pm{0.1}$). At the distance of the $\beta$ Pictoris system ($19.44\pm{}0.05\,\mathrm{pc}$, \citeauthor{Leeuwen2007}~\citeyear{Leeuwen2007}), this corresponds to an absolute K-band magnitude of $M_\mathrm{K} = 12.9\pm{0.1}$.}

{For comparison, the AMES-Cond model predicts a K-band magnitude of $M_K = 12.3\pm0.5$, which is marginally discrepant with regard to our measurement. This means that AMES-Cond overpredicts either the temperature or the radius of the planet. The fact that the Exo-REM atmospheric modelling agrees with the AMES-Cond tracks on the temperature suggests that the AMES-Cond tracks overpredict the radius. Indeed, our atmospheric modelling yields an estimate of $1.2\pm{}0.1\,R_\mathrm{Jup}$ for the radius, which is discrepant with regard to the AMES-Cond prediction ($R=1.4\pm{}0.05\,R_\mathrm{Jup}$)}. 

The position of $\beta$ Pic c in mass/radius and mass/luminosity is shown in Figure~\ref{fig:mass_luminosity}, together with the $18.5\pm{}2.5\,\mathrm{Myr}$ AMES-Cond isochrones, and two synthetic populations of planets formed through core nucleated accretion\footnote{available at \url{https://dace.unige.ch/populationSearch}}. {These two synthetic populations take into account the core-mass effect \citep{Mordasini2013}. They correspond to what is commonly referred to as `warm starts', with their planets much more luminous that in the classical `cold start' scenario \citep{Marley2007}. Compared to the hot-start population, at $M=8\,M_\mathrm{Jup}$, the hot (resp. nominal) core accretion is 30\% less luminous (resp. 50\%) and has radii that are 5\% smaller (resp. 10\%).} For completeness, we carried out the exact same analysis on $\beta$ Pic b, using the published GRAVITY K-band spectrum \citep{GRAVITYCollaboration2020}, and we over-plotted the result in Figure~\ref{fig:mass_luminosity}. Since the dynamical mass of $\beta$ Pic b remains poorly constrained, we used a value of $15.4\pm{}3\,M_\mathrm{Jup}$, based on atmospheric modelling \citep{GRAVITYCollaboration2020}.
The large error bars on the mass of $\beta$ Pic b makes it equally compatible with AMES-Cond and core-accretion, both in terms of  mass/luminosity and mass/radius. This is not the case for  $\beta$ Pic c. Indeed, even if the new GRAVITY measurements presented in this paper are not yet sufficient to confidently reject a hot-start model such as AMES-Cond, the position of the planet in mass/radius, and, to a lesser extent, in mass/luminosity, seems to be in better agreement with a core-accretion model, particularly when the core-mass effect is taken into account. Interestingly, a similar conclusion has been drawn for $\beta$ Pic b \citep{GRAVITYCollaboration2020}  based primarily on the estimation of its C/O abundance ratio.

\section{Conclusion}

{Thanks to this first detection of $\beta$ Pic c, we can now constrain the inclination and the luminosity of the planet. The inclination, in combination with the radial velocities, gives a robust estimate of the mass: $8.2\pm0.8\,M_{\rm Jup}$. On the other hand, the mass of $\beta$ Pic b is not as well-constrained:  $9.0\pm1.6\,M_{\rm Jup}$.}
  
{To match these masses with the hot start scenario, the hot-start AMES-Cond models show that a bigger and brighter $\beta$ Pic c would be needed. Our first conclusion is that the formation of $\beta$ Pic c is not due to gravitational instability but, more likely, to a warm-start core accretion scenario. Our second conclusion is that the mass of $\beta$ Pic b should be revised in the future, once the radial velocity data covers a full orbital period.}

{Our final conclusion is that we are now able to directly observe exoplanets that have been detected by radial velocities. This is an important change in exoplanetary observations because it means we can obtain both the flux and dynamical masses of exoplanets. It also means that we will soon be able to apply direct constrains on the exoplanet formation models.}



\begin{acknowledgements}
This project has received funding from the European Research Council (ERC) under the European Union’s Horizon 2020 research and innovation programme, grant agreement No. 639248 (LITHIUM), 757561 (HiRISE) and 832428 (Origins).
A.A. and P.G. were supported by Funda\c{c}\~{a}o para a Ci\^{e}ncia e a Tecnologia, with grants reference UIDB/00099/2020 and SFRH/BSAB/142940/2018.
\end{acknowledgements}

\begin{appendix}

\section{Observations}
\label{app:obs}

GRAVITY is a fiber-fed interferometer that uses a set of two single-mode fibers for each telescope of the array. One fiber feeds the fringe tracker \citep{Lacour2019}, which is used to compensate for the atmospheric phase variations, and the second fiber feeds the science channel \citep{Pfuhl2014}. The observations of $\beta$ Pictoris c presented in this work were obtained using the on-axis mode, meaning that a beamsplitter was in place to separate the light coming from each telescope. The instrument was also used in dual-field mode, in which the two fibers feeding the fringe tracking and science channels can be positioned independently.

\begin{figure}
  \begin{center}
    \includegraphics[width=\linewidth]{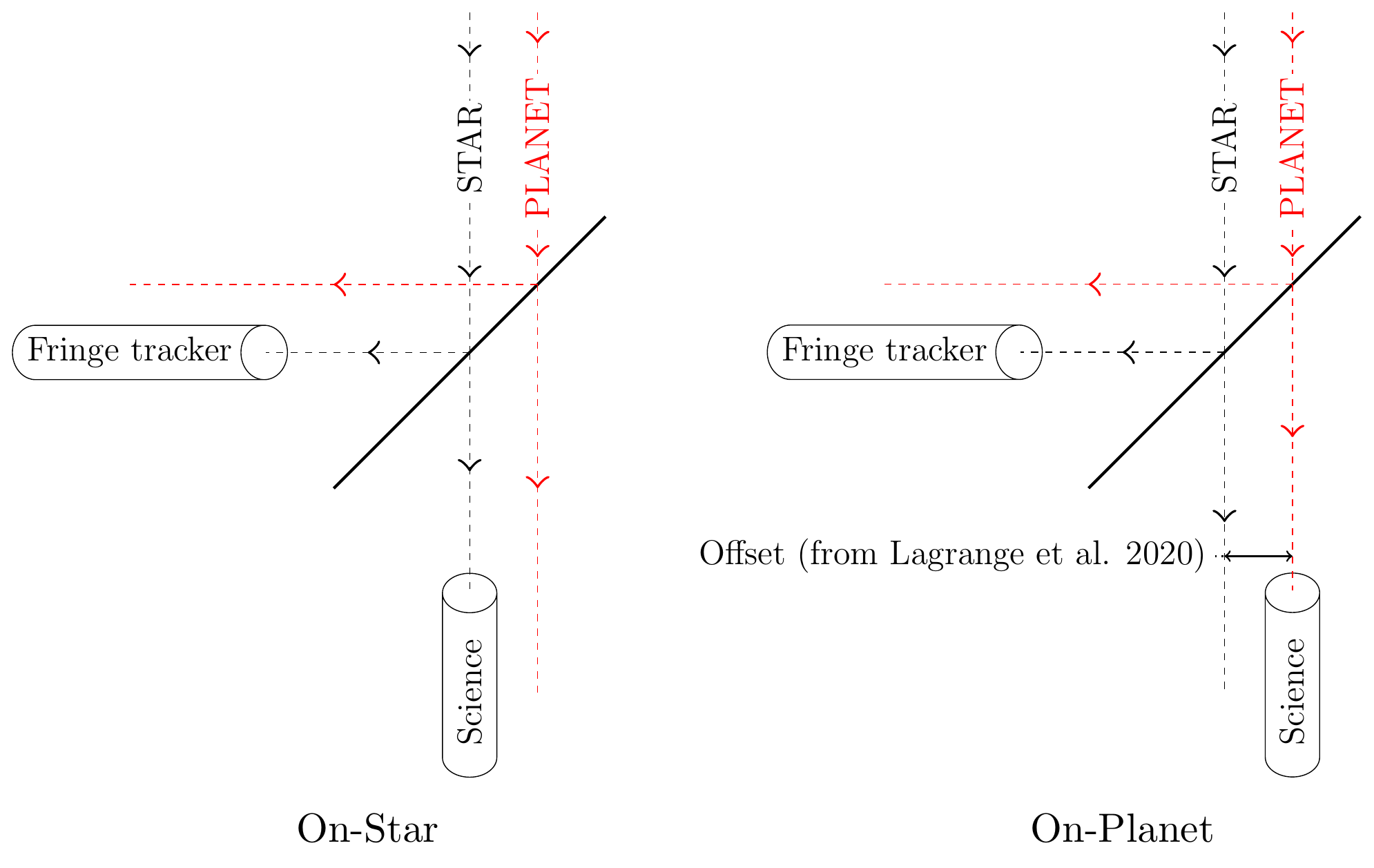}
    \caption{Schematic representation of the positions of the two fibers when observing in on-axis and dual-field mode with VLTI/GRAVITY. In on-axis mode, a beamsplitter is used to separate the incoming light in amplitude. In dual-field, the fringe-tracking fiber and the science fiber can be moved separately, and can be centered on different targets. For observing exoplanets, two types of configuration are used, with the science fiber either centered on the star, for calibration, or centered at the expected location of the planet. The fringe-tracking fiber is always centered on the star, whose light is used as a reference for the phase of the interferometric visibilities.}
    \label{fig:dual_field}
  \end{center}
\end{figure}

The use of the on-axis/dual-field mode is crucial for observing of exoplanets with GRAVITY, as it allows the observer to use a strategy in which the science fiber is alternately set on the central star for phase and amplitude referencing, and at an arbitrary offset position to collect light from the planet (see illustration in Figure~\ref{fig:dual_field}). Since the planet itself cannot be seen on the acquisition camera, its position needs to be known in advance in order to properly position the science fiber. For our observations, we used the most up-to-date predictions for $\beta$ Pic c \citep{Lagrange2020}. The use of single-mode fibers severely limits the field-of-view of the instrument (to about 60~mas), and any error on the position of the fiber results in a flux loss at the output of the combiner (half of the flux is lost for a pointing error of 30 mas). However, a pointing error does not result in a phase error, and therefore does not affect the astrometry or spectrum extraction other than by decreasing the signal-to-noise ratio (S/N). {In practice, any potential flux error is also easily corrected once the exact position of the planet is known (see Appendix~\ref{app:datared}).}

\section{Data reduction}
\label{app:datared}

The theoretical planetary visibility is given as follows:
\begin{equation}
  V_\mathrm{planet}(b, t, \lambda)  = S_\mathrm{planet}(\lambda)\times{}e^{-i\frac{2\pi}{\lambda}\left(\Delta\alpha\times{}U(b, t)+\Delta\delta\times{}V(b, t)\right)}
,\end{equation}
\noindent{}where $S_\mathrm{planet}$ is the spectrum of the planet, $U,V$ are the u-v coordinates of the interferometric baseline $b$, $(\Delta{}\alpha, \Delta\delta)$ the relative astrometry (in RA/dec), $\lambda$ the wavelength, and $t$ the time.

In practice, however, two important factors need to be taken into account: 1) atmospheric and instrumental transmission distort the spectrum, and 2) even when observing on-planet, visibilities are dominated by residual starlight leaking into the fiber. A better representation of the visibility actually observed is given by \citep{GRAVITYCollaboration2020}:
\begin{equation}
  \begin{split}
  V_\mathrm{onplanet}(b, t, \lambda) &= Q(b, t, \lambda)G(b, t, \lambda)V_\mathrm{star}(b, t, \lambda) \\ &+G(b, t, \lambda)V_\mathrm{planet}(b, t, \lambda)
  \label{eq:onplanet}
  \end{split}
,\end{equation}
\noindent{}in which $G$ is a transmission function accounting for both the atmosphere and the instrument, and the $Q(b, t, \lambda)$ are 6th-order polynomial functions of $\lambda$ (one per baseline and per exposure) accounting for the stellar flux.

In parallel, the on-star observations give a measurement of the stellar visibility, also affected the atmospheric and instrumental transmission:
\begin{equation}
  \begin{split}
    V_\mathrm{onstar}(b, t, \lambda) &= G(b, t, \lambda)V_\mathrm{star}(b, t, \lambda) \\ &= G(b, t, \lambda)S_\mathrm{star}(\lambda)
  \end{split}
  \label{eq:onstar}
,\end{equation}
{\noindent{}where the last equality comes from the fact that the star itself is the phase-reference of the observations, and thus $V_\mathrm{star} = S_\mathrm{star}$.}

{Here, if we introduce the contrast spectrum $C = S_\mathrm{planet}/S_\mathrm{star}$, we can get rid of the instrumental and atmospheric transmission:}
\begin{equation}
  \begin{split}
    V_\mathrm{onplanet} &= Q_{b, t}(\lambda)V_\mathrm{onstar}(b, t, \lambda)+\\&C(\lambda)V_\mathrm{onstar}(b, t, \lambda)e^{-i\frac{2\pi}{\lambda}\left(\Delta\alpha{}U(b, t)+\Delta\delta{}V(b, t)\right)}
  \end{split}
  \label{eq:model}
,\end{equation}

Since the $\lambda\rightarrow{}Q(b, t, \lambda)$ are polynomial functions, the model is non-linear only in the two parameters, $\Delta\alpha, \Delta\delta$.

In reality, since the on-star and on-planet data are not acquired simultaneously, Equation~(\ref{eq:onstar}) should be written for $t^*\neq{}t$. This means that the factor $G$ from Equation~(\ref{eq:onplanet}) and (\ref{eq:onstar}) do not cancel out perfectly, and an extra factor $G(t)/G(t^*)$ should appear in Equation~\ref{eq:model}. The two sequences are separated by $10~\mathrm{min}$ at most, and the instrument is stable on such durations. Therefore, the main contribution to this factor $G(t)/G(t^*)$ comes from atmospheric variations. In order to limit the impact of these atmospheric variations on the overall contrast level, {we use the ratio of the fluxes observed by the fringe-tracking fiber during the on-planet observation and the on-star observation} as a proxy for this ratio of $G$. The fringe-tracking combiner works similarly to the science combiner, and acquires data at the same time (though at a higher rate). The main difference is that the fringe-tracking fiber is always centered on the star. Consequently, the ratio of the fluxes observed by the fringe-tracking fiber is a direct estimate of $G(t)/G(t^*)$. Since the fringe-tracking channel has a much lower resolution that the science channel, this only provides a correction of the overall contrast level (integrated in wavelength).

Similarly, a difference in $G$ can occur if the fiber is misaligned with the planet (the fiber is always properly centered on the star, which is observable on the acquisition camera of the instrument). This effect is easily corrected once the correct astrometry of the planet is known, by taking into account the theoretical fiber injection in Equation~\ref{eq:onplanet} and \ref{eq:onstar}. In practice, the initial pointing of the fiber is good and this effect remains small ($<3\%$).

To assess the presence of a planet in the data, we first look for a point-like source in the field-of-view of the fiber, assuming a constant contrast over the K-band: $C(\lambda) = c$. To do so, we switch to a vector representation, and  write $\mathbf{V}_{\mathrm{onplanet}} = (V_\mathrm{onplanet}(b, t, \lambda))_{b, t, \lambda}$ the vector obtained by concatenating all the datapoints (for all the six baselines, all the exposures, and all the wavelength grid points). Similarly, for a given set of polynomials $Q$, and for given values of $c$, $\Delta\alpha$, and $\Delta\delta$, we derive $\mathbf{V}_{\Delta\alpha, \Delta\delta}[Q, c]$, the model for the visibilities described in Equation~\ref{eq:model}. The $\chi^2$ sum is then written:
\begin{equation}
  \begin{split}
    & \chi^2(\Delta\alpha, \Delta\delta, Q, c) = \\ & \left[\mathbf{V}_\mathrm{onplanet} - \mathbf{V}_{\Delta\alpha, \Delta\delta}[Q, c]\right]^T  W^{-1} \left[\mathbf{V}_\mathrm{onplanet} - \mathbf{V}_\mathrm{\Delta\alpha, \Delta\delta}(Q, c)\right]
    \end{split}
,\end{equation}
\noindent{}with $W$ the covariance matrix affecting the projected visibilities $\tilde{\mathbf{V}}$.

This $\chi^2$, minimized over the nuisance parameters, $Q$ and $c$, is related to the likelihood of obtaining the data given the presence of a planet at $\Delta\alpha, \Delta\delta$. The value at $\Delta\alpha = 0$ and $\Delta\delta = 0$ corresponds to the case where no planet is actually injected in the model (the exponential in Equation~\ref{eq:model} is flat) and can be taken as a reference: $\chi^2_\mathrm{no~planet} = \chi^2_\mathrm{planet}(0, 0)$. The values of the $\chi^2$ using a planet model can be compared to this reference, by defining:
\begin{equation}
  z(\Delta\alpha, \Delta\delta) = \chi^2_\mathrm{no~planet} - \chi^2_\mathrm{planet}(\Delta\alpha, \Delta\delta)
,\end{equation}

The quantity $z(\Delta\alpha, \Delta\delta)$ can be understood as a Bayes factor, comparing the likelihood of a model that includes a planet at $(\Delta\alpha, \Delta\delta)$ to the likelihood of a model without any planet. It is also a direct analog of the periodogram power used, for example, in the analysis of radial-velocity data\citep{Scargle1982,Cumming2004}.

The resulting power maps obtained for each of the three nights of $\beta$~Pic~c observation are given in Figure~\ref{fig:chi2Maps}. The astrometry is extracted from these maps by taking the position of the maximum of $z$, and the error bars are obtained by breaking each night into individual exposures so as to estimate the effective standard-deviation from the data themselves.

Once the astrometry is known, Equation~(\ref{eq:model}) can be used to extract the contrast spectrum $C$ \citep{GRAVITYCollaboration2020}. The extraction of the astrometry using $\tilde{\mathbf{V}}_\mathrm{model}$ relies on the implicit assumption that the contrast is constant over the wavelength range. To mitigate this, the whole procedure (astrometry and spectrum extraction) is iterated once more starting with the new contrast spectrum, to check for consistency of results.

\section{Note on multi-planet orbit fitting}
\label{app:orbit}

\begin{figure*}
  \begin{center}
    \includegraphics[width=0.9\linewidth]{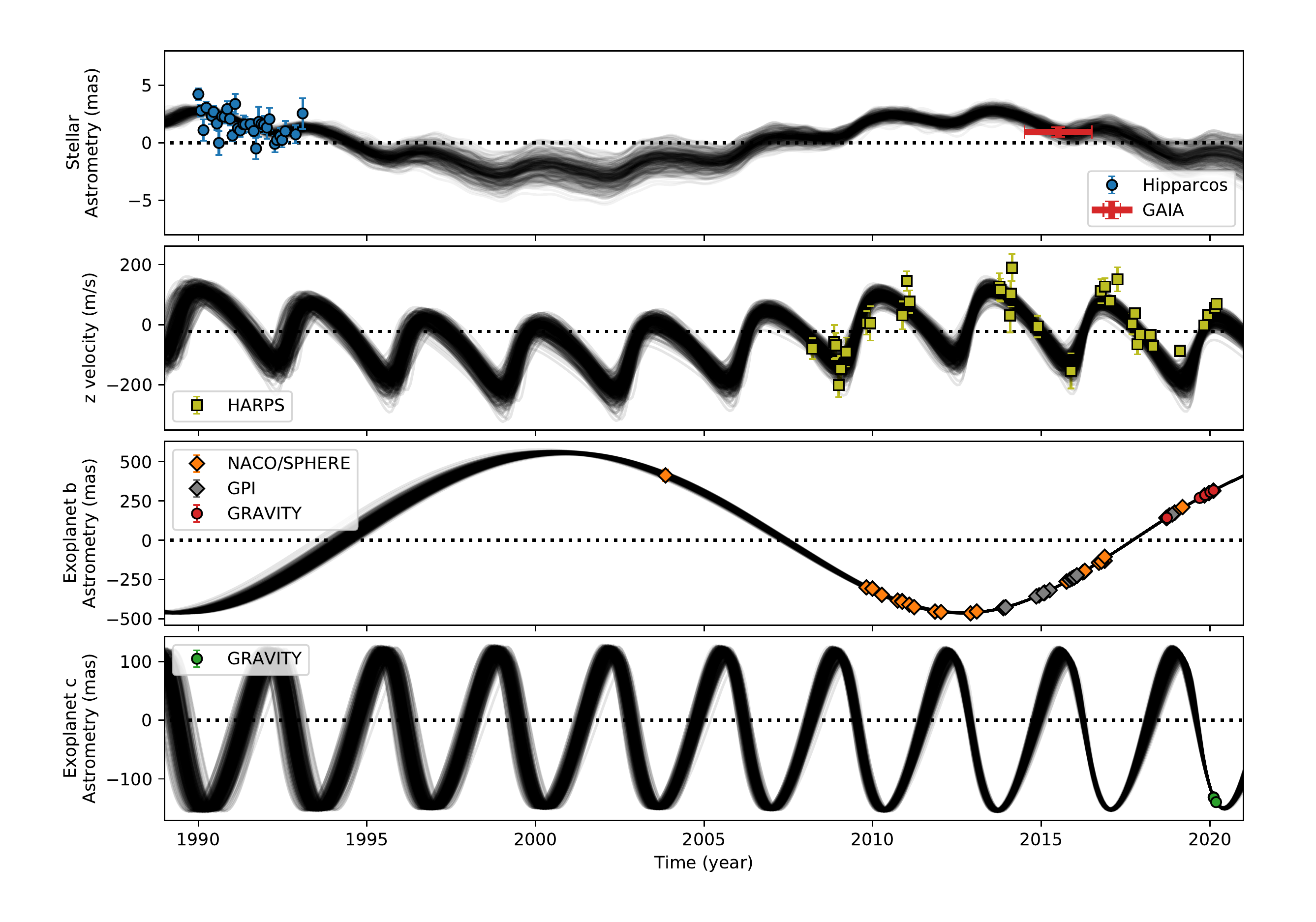}
    \caption{ $\beta$ Pic planetary system: absolute astrometry, radial velocities, and relative astrometry observations.
    Absolute astrometry is showed in the upper panel as deviation from position RA=5:47:17.08346 dec=-51:04:00.1699 at year 1991.25, with a proper motion as in Table~\ref{tab:fit_parameters}.
     Because individual measurements of \emph{Hipparcos} data are 1 dimension only, they are projected in this display along the axis of the $\beta$ Pictoris system, ie. 31.8\,deg.  Radial velocity in the second panel includes the $v_0$ term of -23.09\,m/s (dotted line). The two lower panels show the orbital motion of the planets with respect to the star, also projected on an axis inclined by 31.8\,deg to the East of the North. }
    \label{fig:motion_faceon}
  \end{center}
\end{figure*}

{
For the  MCMC analysis, we used a total of 100 walkers performing 4000 steps each. A random sample of 300 posteriors are plotted in \ref{fig:motion_faceon}. The two lower panels show the trajectory of both planets with respect to the star. Both trajectories are not perfectly Keplerian because each planet influence the position of the star. Therefore, the analysis must be global.
}

The \texttt{orbitize!} \citep{Blunt2020} software  does not permit N-body simulations that account for planet-planet interactions. However, it allows  for the simultaneous modelling of multiple two-body Keplerian orbits.
Each two-body Keplerian orbit is solved in the standard way by reducing it to a one-body problem where we solve for the time evolution of the relative offset of the planet from the star instead of each body's orbit about the system barycenter. As with all imaging astrometry techniques, the GRAVITY measurements are relative separations, so it is convenient to solve orbits in this coordinate system.
However, as each planet orbits,  it perturbs the star which itself orbits around the system barycenter.
 The magnitude of the effect is proportional to the offset of each planet times $M_{\rm planet}/M_{\rm tot}$ where $M_{\rm planet}$ is the mass of the planet and $M_{\rm tot}$ is the total mass of all bodies with separation less than or equal to the separation of the planet. Essentially, the planets cause the star to wobble in its orbit and introduce epicycles in the astrometry of all planets relative to their host star.
 We modelled these mutual perturbations on the relative astrometry in \texttt{orbitize!} and verified it against the \texttt{REBOUND} N-body package that simulates mutual gravitational effects \citep{Rein2012, Rein2015}. We found a maximum disagreement between the packages of 5 $\mu$as over a 15-year span for $\beta$ Pic b and c. This is an order of magnitude smaller than our error bars, so our model of Keplerian orbits with perturbations is a suitable approximation. We are not yet at the point we need to model planet-planet gravitational perturbations that cause the planets' orbital parameters to change in time.

\end{appendix}

\bibliographystyle{aa}
\bibliography{biblio,bibLeati}

\begin{thebibliography}{45}
\expandafter\ifx\csname natexlab\endcsname\relax\def\natexlab#1{#1}\fi

\bibitem[{Baraffe {et~al.}(2003)Baraffe, Chabrier, Barman, Allard, \&
  Hauschildt}]{Baraffe2003}
Baraffe, I., Chabrier, G., Barman, T.~S., Allard, F., \& Hauschildt, P.~H.
  2003, Astronomy \& Astrophysics, 402, 701

\bibitem[{{Beust}(2003)}]{2003A&A...400.1129B}
{Beust}, H. 2003, \aap, 400, 1129

\bibitem[{Blunt {et~al.}(2020)Blunt, Wang, Angelo, Ngo, Cody, Rosa, Graham,
  Hirsch, Nagpal, Nielsen, Pearce, Rice, \& Tejada}]{Blunt2020}
Blunt, S., Wang, J.~J., Angelo, I., {et~al.} 2020, The Astronomical Journal,
  159, 89

\bibitem[{Boss(1997)}]{Boss1997}
Boss, A.~P. 1997, Science, 276, 1836

\bibitem[{Cameron(1978)}]{Cameron1978}
Cameron, A. G.~W. 1978, The moon and the planets, 18, 5

\bibitem[{Charnay {et~al.}(2018)Charnay, B{\'e}zard, Baudino, Bonnefoy,
  Boccaletti, \& Galicher}]{Charnay2018}
Charnay, B., B{\'e}zard, B., Baudino, J.-L., {et~al.} 2018, The Astrophysical
  Journal, 854, 172

\bibitem[{Cumming(2004)}]{Cumming2004}
Cumming, A. 2004, Monthly Notices of the Royal Astronomical Society, 354, 1165

\bibitem[{{GAIA Collaboration} {et~al.}(2018){GAIA Collaboration}, Brown,
  Vallenari, Prusti, de~Bruijne, Babusiaux, {Bailer-Jones}, Biermann, Evans,
  Eyer, Jansen, Jordi, Klioner, Lammers, Lindegren, Luri, Mignard, Panem,
  Pourbaix, Randich, Sartoretti, Siddiqui, Soubiran, van Leeuwen, Walton,
  Arenou, Bastian, Cropper, Drimmel, Katz, Lattanzi, Bakker, Cacciari,
  Casta{\~n}eda, Chaoul, Cheek, Angeli, Fabricius, Guerra, Holl, Masana,
  Messineo, Mowlavi, Nienartowicz, Panuzzo, Portell, Riello, Seabroke, Tanga,
  Th{\'e}venin, {Gracia-Abril}, Comoretto, {Garcia-Reinaldos}, Teyssier,
  Altmann, Andrae, Audard, {Bellas-Velidis}, Benson, Berthier, Blomme, Burgess,
  Busso, Carry, Cellino, Clementini, Clotet, Creevey, Davidson, Ridder,
  Delchambre, Dell'Oro, Ducourant, {Fern{\'a}ndez-Hern{\'a}ndez}, Fouesneau,
  Fr{\'e}mat, Galluccio, {Garc{\'i}a-Torres}, {Gonz{\'a}lez-N{\'u}{\~n}ez},
  {Gonz{\'a}lez-Vidal}, Gosset, Guy, Halbwachs, Hambly, Harrison,
  Hern{\'a}ndez, Hestroffer, Hodgkin, Hutton, Jasniewicz,
  {Jean-Antoine-Piccolo}, Jordan, Korn, {Krone-Martins}, Lanzafame, Lebzelter,
  L{\"o}ffler, Manteiga, Marrese, {Mart{\'i}n-Fleitas}, Moitinho, Mora,
  Muinonen, Osinde, Pancino, Pauwels, Petit, {Recio-Blanco}, Richards,
  Rimoldini, Robin, Sarro, Siopis, Smith, Sozzetti, S{\"u}veges, Torra, van
  Reeven, Abbas, Aramburu, Accart, Aerts, Altavilla, {\'A}lvarez, Alvarez,
  Alves, Anderson, Andrei, Varela, Antiche, Antoja, Arcay, Astraatmadja, Bach,
  Baker, {Balaguer-N{\'u}{\~n}ez}, Balm, Barache, Barata, Barbato, Barblan,
  Barklem, Barrado, Barros, Barstow, Mu{\~n}oz, Bassilana, Becciani,
  Bellazzini, Berihuete, Bertone, Bianchi, Bienaym{\'e}, {Blanco-Cuaresma},
  Boch, Boeche, Bombrun, Borrachero, Bossini, Bouquillon, Bourda, Bragaglia,
  Bramante, Breddels, Bressan, Brouillet, Br{\"u}semeister, Brugaletta,
  Bucciarelli, Burlacu, Busonero, Butkevich, Buzzi, Caffau, Cancelliere,
  Cannizzaro, {Cantat-Gaudin}, Carballo, Carlucci, Carrasco, Casamiquela,
  Castellani, {Castro-Ginard}, Charlot, Chemin, Chiavassa, Cocozza, Costigan,
  Cowell, Crifo, Crosta, Crowley, Cuypers{\textdagger}, Dafonte, Damerdji,
  Dapergolas, David, David, de~Laverny, Luise, March, de~Martino, de~Souza,
  de~Torres, Debosscher, del Pozo, Delbo, Delgado, Delgado, Matteo, Diakite,
  Diener, Distefano, Dolding, Drazinos, Dur{\'a}n, Edvardsson, Enke, Eriksson,
  Esquej, Bontemps, Fabre, Fabrizio, Faigler, Falc{\~a}o, Casas, Federici,
  Fedorets, Fernique, Figueras, Filippi, Findeisen, Fonti, Fraile, Fraser,
  Fr{\'e}zouls, Gai, Galleti, Garabato, {Garc{\'i}a-Sedano}, Garofalo,
  Garralda, Gavel, Gavras, Gerssen, Geyer, Giacobbe, Gilmore, Girona,
  Giuffrida, Glass, Gomes, Granvik, Gueguen, Guerrier, Guiraud,
  {Guti{\'e}rrez-S{\'a}nchez}, Haigron, Hatzidimitriou, Hauser, Haywood,
  Heiter, Helmi, Heu, Hilger, Hobbs, Hofmann, Holland, Huckle, Hypki, Icardi,
  Jan{\ss}en, de~Fombelle, Jonker, Juh{\'a}sz, Julbe, Karampelas, Kewley, Klar,
  Kochoska, Kohley, Kolenberg, Kontizas, Kontizas, Koposov, Kordopatis,
  {Kostrzewa-Rutkowska}, Koubsky, Lambert, Lanza, Lasne, Lavigne, Fustec,
  {Poncin-Lafitte}, Lebreton, Leccia, Leclerc, {Lecoeur-Taibi}, Lenhardt,
  Leroux, Liao, Licata, Lindstr{\o}m, Lister, Livanou, Lobel, L{\'o}pez,
  Managau, Mann, Mantelet, Marchal, Marchant, Marconi, Marinoni,
  Marschalk{\'o}, Marshall, Martino, Marton, Mary, Massari, Matijevi{\v c},
  Mazeh, McMillan, Messina, Michalik, Millar, Molina, Molinaro, Moln{\'a}r,
  Montegriffo, Mor, Morbidelli, Morel, Morris, Mulone, Muraveva, Musella,
  Nelemans, Nicastro, Noval, O'Mullane, Ord{\'e}novic,
  {Ord{\'o}{\~n}ez-Blanco}, Osborne, Pagani, Pagano, Pailler, Palacin,
  Palaversa, Panahi, Pawlak, Piersimoni, Pineau, Plachy, Plum, Poggio,
  Poujoulet, Pr{\v s}a, Pulone, Racero, Ragaini, Rambaux, {Ramos-Lerate},
  Regibo, Reyl{\'e}, Riclet, Ripepi, Riva, Rivard, Rixon, Roegiers, Roelens,
  {Romero-G{\'o}mez}, Rowell, Royer, {Ruiz-Dern}, Sadowski, Sell{\'e}s,
  Sahlmann, Salgado, Salguero, Sanna, {Santana-Ros}, Sarasso, Savietto,
  Schultheis, Sciacca, Segol, Segovia, S{\'e}gransan, Shih, Siltala, Silva,
  Smart, Smith, Solano, Solitro, Sordo, Nieto, Souchay, Spagna, Spoto, Stampa,
  Steele, Steidelm{\"u}ller, Stephenson, Stoev, Suess, Surdej, Szabados,
  {Szegedi-Elek}, Tapiador, Taris, Tauran, Taylor, Teixeira, Terrett,
  Teyssandier, Thuillot, Titarenko, Clotet, Turon, Ulla, Utrilla, Uzzi,
  Vaillant, Valentini, Valette, van Elteren, Hemelryck, van Leeuwen, Vaschetto,
  Vecchiato, Veljanoski, Viala, Vicente, Vogt, von Essen, Voss, Votruba,
  Voutsinas, Walmsley, Weiler, Wertz, Wevers, Wyrzykowski, Yoldas, {\v Z}erjal,
  Ziaeepour, Zorec, Zschocke, Zucker, Zurbach, \&
  Zwitter}]{GAIACollaboration2018}
{GAIA Collaboration}, Brown, A. G.~A., Vallenari, A., {et~al.} 2018, Astronomy
  \& Astrophysics, 616, A1

\bibitem[{{GRAVITY Collaboration} {et~al.}(2017){GRAVITY Collaboration},
  Abuter, Accardo, Amorim, Anugu, {\'A}vila, Azouaoui, Benisty, Berger, Blind,
  Bonnet, Bourget, Brandner, Brast, Buron, Burtscher, Cassaing, Chapron,
  Choquet, Cl{\'e}net, Collin, du~Foresto, de~Wit, de~Zeeuw, Deen,
  {Delplancke-Str{\"o}bele}, Dembet, Derie, Dexter, Duvert, Ebert, Eckart,
  Eisenhauer, Esselborn, F{\'e}dou, Finger, Garcia, Dabo, Lopez, Gendron,
  Genzel, Gillessen, Gonte, Gordo, Grould, Gr{\"o}zinger, Guieu, Haguenauer,
  Hans, Haubois, Haug, Haussmann, Henning, Hippler, Horrobin, Huber, Hubert,
  Hubin, Hummel, Jakob, Janssen, Jochum, Jocou, Kaufer, Kellner, Kendrew, Kern,
  Kervella, Kiekebusch, Klein, Kok, Kolb, Kulas, Lacour, Lapeyr{\`e}re,
  Lazareff, Bouquin, L{\`e}na, Lenzen, L{\'e}v{\^e}que, Lippa, Magnard,
  Mehrgan, Mellein, M{\'e}rand, {Moreno-Ventas}, Moulin, M{\"u}ller,
  M{\"u}ller, Neumann, Oberti, Ott, Pallanca, Panduro, Pasquini, Paumard,
  Percheron, Perraut, Perrin, Pfl{\"u}ger, Pfuhl, Duc, Plewa, Popovic, Rabien,
  Ram{\'i}rez, Ramos, Rau, Riquelme, Rohloff, Rousset, {Sanchez-Bermudez},
  Scheithauer, Sch{\"o}ller, Schuhler, Spyromilio, Straubmeier, Sturm, Suarez,
  Tristram, Ventura, Vincent, Waisberg, Wank, Weber, Wieprecht, Wiest,
  Wiezorrek, Wittkowski, Woillez, Wolff, Yazici, Ziegler, \&
  Zins}]{GRAVITYCollaboration2017}
{GRAVITY Collaboration}, Abuter, R., Accardo, M., {et~al.} 2017, Astronomy \&
  Astrophysics, 602, A94

\bibitem[{{GRAVITY Collaboration} {et~al.}(2019){GRAVITY Collaboration},
  Lacour, Nowak, Wang, Pfuhl, Eisenhauer, Abuter, Amorim, Anugu, Benisty,
  Berger, Beust, Blind, Bonnefoy, Bonnet, Bourget, Brandner, Buron, Collin,
  Charnay, Chapron, Cl{\'e}net, du~Foresto, de~Zeeuw, Deen, Dembet, Dexter,
  Duvert, Eckart, Schreiber, F{\'e}dou, Garcia, Lopez, Gao, Gendron, Genzel,
  Gillessen, Gordo, Greenbaum, Habibi, Haubois, Hau{\ss}mann, Henning, Hippler,
  Horrobin, Hubert, Rosales, Jocou, Kendrew, Kervella, Kolb, Lagrange,
  Lapeyr{\`e}re, Bouquin, L{\'e}na, Lippa, Lenzen, Maire, Molli{\`e}re, Ott,
  Paumard, Perraut, Perrin, Pueyo, Rabien, Ram{\'i}rez, Rau,
  {Rodr{\'i}guez-Coira}, Rousset, {Sanchez-Bermudez}, Scheithauer, Schuhler,
  Straub, Straubmeier, Sturm, Tacconi, Vincent, van Dishoeck, von Fellenberg,
  Wank, Waisberg, Widmann, Wieprecht, Wiest, Wiezorrek, Woillez, Yazici,
  Ziegler, \& Zins}]{GRAVITYCollaboration2019}
{GRAVITY Collaboration}, Lacour, S., Nowak, M., {et~al.} 2019, Astronomy \&
  Astrophysics, 623, L11

\bibitem[{{GRAVITY Collaboration} {et~al.}(2020){GRAVITY Collaboration}, Nowak,
  Lacour, Molli{\`e}re, Wang, Charnay, van Dishoeck, Abuter, Amorim, Berger,
  Beust, Bonnefoy, Bonnet, Brandner, Buron, Cantalloube, Collin, Chapron,
  Cl{\'e}net, du~Foresto, de~Zeeuw, Dembet, Dexter, Duvert, Eckart, Eisenhauer,
  Schreiber, F{\'e}dou, Lopez, Gao, Gendron, Genzel, Gillessen, Hau{\ss}mann,
  Henning, Hippler, Hubert, Jocou, Kervella, Lagrange, Lapeyr{\`e}re, Bouquin,
  L{\'e}na, Maire, Ott, Paumard, {C. Paladini}, Perraut, Perrin, Pueyo, Pfuhl,
  Rabien, Rau, {Rodr{\'i}guez-Coira}, Rousset, Scheithauer, Shangguan, Straub,
  Straubmeier, Sturm, Tacconi, Vincent, Widmann, Wieprecht, Wiezorrek, Woillez,
  Yazici, \& Ziegler}]{GRAVITYCollaboration2020}
{GRAVITY Collaboration}, Nowak, M., Lacour, S., {et~al.} 2020, Astronomy \&
  Astrophysics, 633, A110

\bibitem[{Hauschildt {et~al.}(1999)Hauschildt, Allard, \&
  Baron}]{Hauschildt1999}
Hauschildt, P.~H., Allard, F., \& Baron, E. 1999, The Astrophysical Journal,
  512, 377

\bibitem[{Helling {et~al.}(2008)Helling, Dehn, Woitke, \&
  Hauschildt}]{Helling2008}
Helling, C., Dehn, M., Woitke, P., \& Hauschildt, P.~H. 2008, The Astrophysical
  Journal Letters, 675, L105

\bibitem[{Helling \& Woitke(2006)}]{Helling2006}
Helling, C. \& Woitke, P. 2006, Astronomy \& Astrophysics, 455, 325

\bibitem[{Kervella {et~al.}(2019)Kervella, Arenou, Mignard, \&
  Th{\'e}venin}]{Kervella2019}
Kervella, P., Arenou, F., Mignard, F., \& Th{\'e}venin, F. 2019, Astronomy \&
  Astrophysics, 623, A72

\bibitem[{Kraus {et~al.}(2020)Kraus, LeBouquin, Kreplin, Davies, Hone, Monnier,
  Gardner, Kennedy, \& Hinkley}]{Kraus2020}
Kraus, S., LeBouquin, J.-B., Kreplin, A., {et~al.} 2020, arXiv:2006.10784
  [astro-ph] [\eprint[arXiv]{2006.10784}]

\bibitem[{Lacour {et~al.}(2019)Lacour, Dembet, Abuter, F{\'e}dou, Perrin,
  Choquet, Pfuhl, Eisenhauer, Woillez, Cassaing, Wieprecht, Ott, Wiezorrek,
  Tristram, Wolff, Ram{\'i}rez, Haubois, Perraut, Straubmeier, Brandner, \&
  Amorim}]{Lacour2019}
Lacour, S., Dembet, R., Abuter, R., {et~al.} 2019, Astronomy \& Astrophysics,
  624, A99

\bibitem[{Lagrange \& al.(in preparation)}]{Lagrange2020}
Lagrange, A.-M. \& al. in preparation, A \& A

\bibitem[{Lagrange {et~al.}(2019)Lagrange, Meunier, Rubini, Keppler, Galland,
  Chapellier, Michel, Balona, Beust, Guillot, Grandjean, Borgniet,
  M{\'e}karnia, Wilson, Kiefer, Bonnefoy, {Lillo-Box}, Pantoja, Jones,
  Iglesias, Rodet, Diaz, Zapata, Abe, \& Schmider}]{Lagrange2019}
Lagrange, A.-M., Meunier, N., Rubini, P., {et~al.} 2019, Nature Astronomy, 3,
  1135

\bibitem[{Lapeyrere {et~al.}(2014)Lapeyrere, Kervella, Lacour, Azouaoui,
  {Garcia-Dabo}, Perrin, Eisenhauer, Perraut, Straubmeier, Amorim, \&
  Brandner}]{Lapeyrere2014}
Lapeyrere, V., Kervella, P., Lacour, S., {et~al.} 2014, in Optical and
  {{Infrared Interferometry IV}}, Vol. 9146 ({International Society for Optics
  and Photonics}), 91462D

\bibitem[{Lissauer \& Stevenson(2007)}]{2007prpl.conf..591L}
Lissauer, J.~J. \& Stevenson, D.~J. 2007, in Protostars and Planets v, ed.
  B.~Reipurth, D.~Jewitt, \& K.~Keil, 591

\bibitem[{Marleau \& Cumming(2014)}]{Marleau2014}
Marleau, G.-D. \& Cumming, A. 2014, Monthly Notices of the Royal Astronomical
  Society, 437, 1378

\bibitem[{Marleau {et~al.}(2017)Marleau, Klahr, Kuiper, \&
  Mordasini}]{Marleau2017}
Marleau, G.-D., Klahr, H., Kuiper, R., \& Mordasini, C. 2017, The Astrophysical
  Journal, 836, 221

\bibitem[{Marleau {et~al.}(2019)Marleau, Mordasini, \& Kuiper}]{Marleau2019}
Marleau, G.-D., Mordasini, C., \& Kuiper, R. 2019, The Astrophysical Journal,
  881, 144

\bibitem[{Marley {et~al.}(2007)Marley, Fortney, Hubickyj, Bodenheimer, \&
  Lissauer}]{Marley2007}
Marley, M.~S., Fortney, J.~J., Hubickyj, O., Bodenheimer, P., \& Lissauer,
  J.~J. 2007, The Astrophysical Journal, 655, 541

\bibitem[{Miret-Roig \& al.(2020)}]{MiretRoig2020}
Miret-Roig, N. \& al. 2020, A \& A

\bibitem[{Molli{\`e}re {et~al.}(2020)Molli{\`e}re, Stolker, Lacour, Otten,
  Shangguan, Charnay, Molyarova, Nowak, Henning, Semenov, {van Dishoeck},
  Eisenhauer, Garcia, Greenbaum, Kervella, Kreidberg, Maire, Nasedkin, \& {et
  al.}}]{Molliere2020}
Molli{\`e}re, P., Stolker, T., Lacour, S., {et~al.} 2020, Astronomy \&
  Astrophysics

\bibitem[{Mordasini(2013)}]{Mordasini2013}
Mordasini, C. 2013, Astronomy \& Astrophysics, 558, A113

\bibitem[{Mordasini {et~al.}(2017)Mordasini, Marleau, \&
  Molli{\`e}re}]{Mordasini2017}
Mordasini, C., Marleau, G.-D., \& Molli{\`e}re, P. 2017, Astronomy \&
  Astrophysics, 608, A72

\bibitem[{Nesvorn{\'y}(2018)}]{Nesvorny2018}
Nesvorn{\'y}, D. 2018, Annual Review of Astronomy and Astrophysics, 56, 137

\bibitem[{Nielsen {et~al.}(2020)Nielsen, Rosa, Wang, Sahlmann, Kalas,
  Duch{\^e}ne, Rameau, Marley, Saumon, Macintosh, {Millar-Blanchaer}, Nguyen,
  Ammons, Bailey, Barman, Bulger, Chilcote, Cotten, Doyon, Esposito,
  Fitzgerald, Follette, Gerard, Goodsell, Graham, Greenbaum, Hibon, Hung,
  Ingraham, Konopacky, Larkin, Maire, Marchis, Marois, Metchev, Oppenheimer,
  Palmer, Patience, Perrin, Poyneer, Pueyo, Rajan, Rantakyr{\"o}, Ruffio,
  Savransky, Schneider, Sivaramakrishnan, Song, Soummer, Thomas, Wallace,
  {Ward-Duong}, Wiktorowicz, \& Wolff}]{Nielsen2020}
Nielsen, E.~L., Rosa, R. J.~D., Wang, J.~J., {et~al.} 2020, The Astronomical
  Journal, 159, 71

\bibitem[{{Petrovich}(2015)}]{2015ApJ...808..120P}
{Petrovich}, C. 2015, \apj, 808, 120

\bibitem[{Pfuhl {et~al.}(2014)Pfuhl, Haug, Eisenhauer, Kellner, Haussmann,
  Perrin, Gillessen, Straubmeier, Ott, {Rousselet-Perraut}, Amorim, Lippa,
  Janssen, Brandner, Kok, Blind, Burtscher, Sturm, Wieprecht, Schoeller, Weber,
  Hans, \& Huber}]{Pfuhl2014}
Pfuhl, O., Haug, M., Eisenhauer, F., {et~al.} 2014, in Optical and {{Infrared
  Interferometry IV}}, Vol. 9146 ({International Society for Optics and
  Photonics}), 914623

\bibitem[{Pollack {et~al.}(1996)Pollack, Hubickyj, Bodenheimer, Lissauer,
  Podolak, \& Greenzweig}]{Pollack1996}
Pollack, J.~B., Hubickyj, O., Bodenheimer, P., {et~al.} 1996, Icarus, 124, 62

\bibitem[{Rafikov(2005)}]{Rafikov2005}
Rafikov, R.~R. 2005, The Astrophysical Journal Letters, 621, L69

\bibitem[{Rein \& Liu(2012)}]{Rein2012}
Rein, H. \& Liu, S.-F. 2012, Astronomy \& Astrophysics, 537, A128

\bibitem[{Rein \& Spiegel(2015)}]{Rein2015}
Rein, H. \& Spiegel, D.~S. 2015, Monthly Notices of the Royal Astronomical
  Society, 446, 1424

\bibitem[{Scargle(1982)}]{Scargle1982}
Scargle, J.~D. 1982, The Astrophysical Journal, 263, 835

\bibitem[{Snellen \& Brown(2018)}]{Snellen2018}
Snellen, I. a.~G. \& Brown, A. G.~A. 2018, Nature Astronomy, 2, 883

\bibitem[{Stolker {et~al.}(2020)Stolker, Quanz, Todorov, K{\"u}hn,
  Molli{\`e}re, Meyer, Currie, Daemgen, \& Lavie}]{Stolker2020}
Stolker, T., Quanz, S.~P., Todorov, K.~O., {et~al.} 2020, Astronomy \&
  Astrophysics, 635, A182

\bibitem[{van~der Bliek {et~al.}(1996)van~der Bliek, Manfroid, \&
  Bouchet}]{Bliek1996}
van~der Bliek, N.~S., Manfroid, J., \& Bouchet, P. 1996, Astronomy and
  Astrophysics Supplement Series, 119, 547

\bibitem[{van Leeuwen(2007)}]{Leeuwen2007}
van Leeuwen, F. 2007, Astronomy \& Astrophysics, 474, 653

\bibitem[{Woitke \& Helling(2003)}]{Woitke2003}
Woitke, P. \& Helling, C. 2003, Astronomy \& Astrophysics, 399, 297

\bibitem[{Zorec \& Royer(2012)}]{Zorec2012}
Zorec, J. \& Royer, F. 2012, Astronomy \& Astrophysics, 537, A120

\bibitem[{Zwintz {et~al.}(2019)Zwintz, Reese, Neiner, Pigulski, Kuschnig,
  M{\"u}llner, Zieba, Abe, Guillot, Handler, Kenworthy, Stuik, Moffat,
  Popowicz, Rucinski, Wade, Weiss, Bailey, Crawford, Ireland, Lomberg, Mamajek,
  Mellon, \& Talens}]{Zwintz2019}
Zwintz, K., Reese, D.~R., Neiner, C., {et~al.} 2019, Astronomy \& Astrophysics,
  627, A28

\end{thebibliography}

\end{document}